\documentclass[a4paper,fleqn,usenatbib,useAMS]{mnras}

\usepackage{graphicx}	
\usepackage{amsmath}	
\usepackage{amssymb}	
\usepackage{multicol}        
\usepackage{multirow}
\usepackage{bm}		
\usepackage{pdflscape}	

\usepackage[T1]{fontenc}
\usepackage{ae,aecompl}
\usepackage{mathptmx}
\usepackage{longtable}
\usepackage{array}
\usepackage{booktabs}
\let\oldAA\AA
\renewcommand{\AA}{\text{\normalfont\oldAA}}

\usepackage{color}

\title[Internal structures of MaNGA early-type galaxies]{SDSS-IV MaNGA: Internal mass distributions and orbital structures of early-type galaxies and their dependence on environment}

\author[Y. Jin et al.]
{Yunpeng Jin$^{1,2}$\thanks{E-mail:jyp199333@163.com}, Ling Zhu$^{3}$, R. J. Long$^{4,1,5}$, Shude Mao$^{4,1}$, Lan Wang$^{1}$, Glenn van de Ven$^{6,7}$\\\\
$^{1}$National Astronomical Observatories, Chinese Academy of Sciences, 20A Datun Road, Chaoyang District, Beijing 100101, China\\
$^{2}$University of Chinese Academy of Sciences, Beijing 100049, China\\
$^{3}$Shanghai Astronomical Observatory, Chinese Academy of Sciences, 80 Nandan Road, Shanghai 200030, China\\
$^{4}$Department of Astronomy and Tsinghua Centre for Astrophysics, Tsinghua University, Beijing 100084, China\\
$^{5}$Jodrell Bank Centre for Astrophysics, Department of Physics and Astronomy, The University of Manchester, Oxford Road, Manchester M13 9PL, UK\\
$^{6}$Department of Astrophysics, University of Vienna, T\"urkenschanzstrasse 17, 1180 Vienna, Austria\\
$^{7}$European Southern Observatory, Karl-Schwarzschild-Str. 2, 85748 Garching b. M\"unchen, Germany
}

\date{Accepted XXX. Received YYY; in original form ZZZ}
\pubyear{2019}

\begin{document}
\label{firstpage}
\pagerange{\pageref{firstpage}--\pageref{lastpage}}
\maketitle

\begin{abstract}
In our earlier 2019 paper, we evaluated the reliability of Schwarzschild's orbit-superposition dynamical modelling method in estimating the internal mass distribution, intrinsic stellar shapes and orbit distributions of early-type galaxies (ETGs) taken from the Illustris cosmological simulation. We now apply the same techniques to galaxies taken from the integral-field survey Mapping Nearby Galaxies with APO (MaNGA), using a sample of 149 ETGs in the mass range of $10^{9.90}\sim10^{11.80} M_{\odot}$ and made up of 105 central and 44 satellite galaxies. We find that low-mass ETGs with $\log(M_*/M_{\odot})\textless11.1$ have an average dark matter fraction of $\sim0.2$ within one effective radius $R_{\rm e}$, tend to be oblate-like, and are dominated by rotation about their minor axis. High-mass ETGs with $\log(M_*/M_{\odot})\textgreater11.1$ have an average dark matter fraction of $\sim0.4$ within one effective radius $R_{\rm e}$, tend to be prolate-like, and are dominated by rotation about their major axis and by centrophilic orbits. The changes of internal structures within one $R_{\rm e}$ are dominated by the total stellar mass of the individual galaxies. We find no differences of internal structures between central and satellite ETGs for the same stellar masses. However, for similar stellar mass and colour distributions, we find that ETGs more prolate-like, or with more hot orbits,  tend to have higher close neighbour counts at $r_p\sim40$ kpc.
\end{abstract}

\begin{keywords}
galaxies: elliptical and lenticular, cD -- galaxies: kinematics and dynamics -- galaxies: structure -- galaxies: fundamental parameters
\end{keywords}

\section{Introduction}
\label{sec1}
In the currently accepted paradigm of cosmological structure formation, it is widely believed that galaxies form and evolve in cold dark matter haloes. In the $\Lambda$CDM framework, galaxies undergo hierarchical evolution, with smaller galaxies merging to form larger galaxies. The merger history of a galaxy is thought to be one of the major factors that determines the internal kinematic structures of galaxies (e.g., \citealp{White1979,Fall1980,Park2019}). However, galactic structures can also evolve via secular processes, perhaps due to disk internal instabilities, or heating from bars or spiral arms (see, for example,
 \citealp{Minchev2006,Saha2010}).

Statistical studies show that structural properties of galaxies vary as a function of galaxy stellar mass (e.g., bulge-to-total mass fraction, \citealp{Weinzirl2009}; size and S\'ersic index, \citealp{Wuyts2011}; Hubble type, \citealp{Bernardi2010} and \citealp{Calvi2012}; velocity dispersion $\sigma_{\rm e}$ and size, \citealp{Cappellari2013b}). The environment of a galaxy also plays an important role in creating structures via actions such as tidal stripping or ram pressure. For example, a bar could be induced by tidal forces in a cluster environment \citep{Lokas2016}. There are a number of ways of quantifying the effect of environment on galaxies. Galaxy groups have been separated into central and satellite galaxies (e.g., \citealp{Zheng2005,Berlind2006,Yang2007,Yang2008}) based on the halo occupation distribution. Unsurprisingly, central and satellite galaxies have both similarities and differences. Satellites are older and metal richer than centrals of the same stellar mass \citep{Pasquali2010}, while they both follow a qualitatively similar gas-phase metallicity \citep{Pasquali2012}. Late-type satellites are found to be redder, smaller and more concentrated than late-type centrals for a given stellar mass \citep{Weinmann2009}. Such differences do not exist between early-type centrals and satellites \citep{Huertas-Company2013}. There is no significant difference of bugle-to-total mass ratio B/T between centrals and satellites \citep{Bluck2019}. Quite apart from this separation into centrals and satellites, a galaxy's local density environment can affect its properties, such as morphology \citep{Postman1984,Dressler1997,Goto2003,Tanaka2004}, or luminosity and colour \citep{Hogg2003,Blanton2005a,Deng2009}. A galaxy's local density environment can be quantified by the number of neighbours $N_{\rm neighbour}$ within a certain projected radius $r_p$ \citep{Li2008}. The number of neighbours seems to be related to a galaxy's internal structures, with, for example, pseudo-bulge and pure-disc galaxies showing a strong excess in close neighbour counts when compared to control galaxies with similar stellar masses \citep{Wang2019}.

In recent years, Integral Field Unit (IFU) surveys such as SAURON \citep{Bacon2001}, $\rm ATLAS^{3D}$ \citep{Cappellari2011}, CALIFA \citep{Sanchez2012}, SAMI \citep{Bryant2015}, MaNGA \citep{Bundy2015,Blanton2017} and MUSE \citep{Bacon2017} provide us with much data about galaxies potentially allowing their structure, formation and evolution to be investigated. The MaNGA survey (Mapping Nearby Galaxies at Apache Point Observatory), one of three core programmes in SDSS-IV (the fourth-generation Sloan Digital Sky Survey), aims to observe a sample of 10,000 nearby galaxies, and is the survey from which we take our galaxy data.

Using data from IFU surveys, we are able to investigate the internal kinematics of galaxies. \citet{Greene2018} analyzed the two-dimensional stellar kinematic maps of MaNGA galaxies, and found that there is no residual differences in angular momentum content $\lambda_R$ \citep{Emsellem2011} between central and satellite early-type galaxies when carefully matching the stellar mass distributions. However, $\lambda_R$ is directly calculated from kinematic maps on the observational plane and can only incorporate line-of-sight kinematics. Our earlier paper \citet{Jin2019}, as a precursor to this paper, evaluated the capabilities of Schwarzschild orbit-superposition technique \citep{Schwarzschild1979} using data from the Illustris cosmological simulation \citep{Vogelsberger2014a,Vogelsberger2014b,Genel2014,Nelson2015}, and we continue with Schwarzschild's method and the same implementation here.
Stellar dynamical modelling tools such as Schwarzschild's method are only able to provide partial insight into the three-dimensional internal kinematic structures underlying the line-of-sight kinematics. Given the limitations of the instrumentation currently available for observing external galaxies and for reasons of deprojection non-uniqueness, these tools can not identify the true three-dimensional structures of a galaxy but can at least give an indication as to what they might be like.

Early-type galaxies (ETGs) show complicated structures in their kinematic maps, but can be split on their kinematics into two classes, fast rotators and slow rotators. According to \citet{Cappellari2016}, this split into two classes also indicates two major channels of galaxy formation. Fast-rotator ETGs start as star forming disks and evolve through a set of processes dominated by gas accretion, bulge growth and quenching. By comparison, slow-rotator ETGs assemble near the centre of massive haloes via intense star formation at high redshift, and evolve from a set of processes dominated by gas poor mergers. Environment is believed to play an important role in the quenching of star formation \citep{Peng2010}. However, the extent to which environmental processes could also affect internal galaxy structures is unclear.

For the research documented here, we have focussed on ETGs only, and our intention is to examine the relationships between internal structures, evolution and environments. To this end, our specific objectives, using a sample of central and satellite ETGs taken from MaNGA, are
\begin{enumerate}
\item to model the galaxies individually using Schwarzschild's method and determine their mass distributions, intrinsic stellar shapes and internal orbit distributions,
\item to examine statistically the differences and similarities between central and satellite ETGs, and
\item to assess the role of the environment in the galaxies' evolution.
\end{enumerate}
Our sample of 149 galaxies is to date the largest sample of ETGs modelled using Schwarzschild's method. \citet{Zhu2018b} used the same
triaxial Schwarzschild implementation \citep{RvdB2008} in their modelling of 300 CALIFA galaxies but over 70 percent of their galaxy sample were LTGs. We use cosmological parameters from \citet{Planck2014}: $H_0=67.8\ \rm km\cdot s^{-1}\cdot Mpc^{-1}$, $\Omega_{\rm \Lambda}=0.692$ and $\Omega_{\rm m}=0.308$.

The structure of the paper is as follows. In $\S$~\ref{sec2}, we describe at a high level the approach we take which is based strongly on our experiences of having evaluated Schwarzschild's method with simulation data \citep{Jin2019}. In $\S$~\ref{sec3}, our galaxy sample selection criteria and the MaNGA data we use are presented. Low level technical details about our modelling are given in $\S$~\ref{sec4}. In $\S$~\ref{sec5}, we analyze the model results for an example galaxy. The statistical results from modelling our galaxy sample are shown in $\S$~\ref{sec6}, with the local environment impact being considered in $\S$~\ref{sec7}. We discuss our findings in $\S$~\ref{sec8}, and conclude our investigation in $\S$~\ref{sec9}.

\section{Approach}
\label{sec2}
As indicated in $\S$~\ref{sec1}, we are modelling our individual galaxies using Schwarzschild's method, and we use the implementation produced by \citet{RvdB2008} which allows us to model triaxial stellar components. A systematic evaluation of this implementation, using data from the Illustris simulation, has already been presented in our companion paper \citep{Jin2019}. In the evaluation, nine early-type simulated galaxies with a range of triaxiality were selected and used to create mock IFU data with similar data quality to that of MaNGA. Five line-of-sight projections of each galaxy were made giving a total of 45 galaxies to be modelled. We assessed estimates of the mass profiles including both stellar and dark matter components, galaxy intrinsic stellar shapes, and orbit circularities. The masses and shapes were reasonably close to the true values of the simulated galaxies while the 3D orbit circularities were plausible. The quantified model biases and uncertainties are used in this investigation (see $\S$~\ref{sec6.1} for details).

In this paper, we set up our Schwarzschild models for our MaNGA galaxies in the same way as described in \citet{Jin2019}. We refer the readers to \citet{RvdB2008} and \citet{Jin2019} for fuller details on triaxial Schwarzschild modelling. The galaxy gravitational potentials in our modelling are generated in three parts, as in \citet{Jin2019}: a triaxial stellar component, whose mass is calculated using the Multi-Gaussian Expansion (MGE) formalism \citep{Emsellem1994,Cappellari2002}; a dark matter halo, which is assumed to follow the spherical Navarro-Frenk-White (NFW) profile \citep{Navarro1996}; and a central super massive black hole which is treated as a point source. There are six free parameters describing the
overall potential which must be determined. As in our earlier paper, we run a grid based parameter search to determine the best-fitting parameters for each galaxy. The search itself requires 1000 to 2000 separate Schwarzschild models per galaxy to be run.

Previous research using the \citet{RvdB2008} implementation indicates that it has been well-utilised over a 10 year period prior to our work. Using mock galaxies based on theoretical Abel models \citep{Dejonghe1991,Mathieu1999}, \citet{vdV2008} estimated internal orbital structures and \citet{RvdB2009} investigated how well the model could match the true intrinsic stellar shapes. \citet{RvdB2010} modelled two individual galaxies NGC 3379 and M32 and obtained their black hole masses. \citet{Zhu2018a} investigated the orbits of CALIFA spiral galaxies, and \citet{Zhu2018b} evaluated how well the model described the true circularity distributions of over 100 mock galaxies from different simulations, and also studied the statistical circularity distributions for 300 CALIFA galaxies.

Having obtained our best-fitting Schwarzschild models, we need a stellar mass for each galaxy that can be used to compare our results with other results from the literature. Unfortunately, many of these literature results are framed not using dynamical stellar masses but masses obtained from population synthesis techniques. In addition results from investigating orbits and orbit structures are quoted using stellar masses only and not the total mass including dark matter. In this paper, we choose to discard the dynamical stellar masses resulting from using Schwarzschild method and substitute instead stellar masses determined by stellar population synthesis. This approach is not completely satisfactory and we will return to it in the discussion in $\S$~\ref{sec8}.

In assessing the role of environment in the evolution of our galaxies, we do so by considering how many other galaxies are in the vicinity of each galaxy. We utilise galaxy neighbour counts \citep{Li2008}, and study whether or not there is any variation in the internal properties of our galaxies with differing neighbour counts.

\section{Data}
\label{sec3}
\subsection{MaNGA}
\label{sec3.1}
The galaxy selection criteria of MaNGA depend on i-band absolute magnitude and ensure that all galaxies have similar angular size so that the IFUs can cover 1.5 to 2.5 effective radii per galaxy \citep{Wake2017}. MaNGA uses the 2.5m Sloan Telescope \citep{Gunn2006} and employs 17 hexagonal fibre bundles that vary in diameter from 12$\arcsec$ (19 fibres) to 32$\arcsec$ (127 fibres) \citep{Drory2015}. The twin multi-object fibre spectrographs cover a wide range of wavelengths, from 360 $\AA$ to 1030 $\AA$ \citep{Smee2013}. Other important information about MaNGA includes the observing strategy in \citet{Law2015}, the survey design in \citet{Yan2016a}, the spectrophotometry calibration technique in \citet{Yan2016b}, and the data reduction pipeline (DRP) in \citet{Law2016}.

\subsection{Sample Selection}
\label{sec3.2}
\begin{figure*}
\begin{centering}
	\includegraphics[width=14cm]{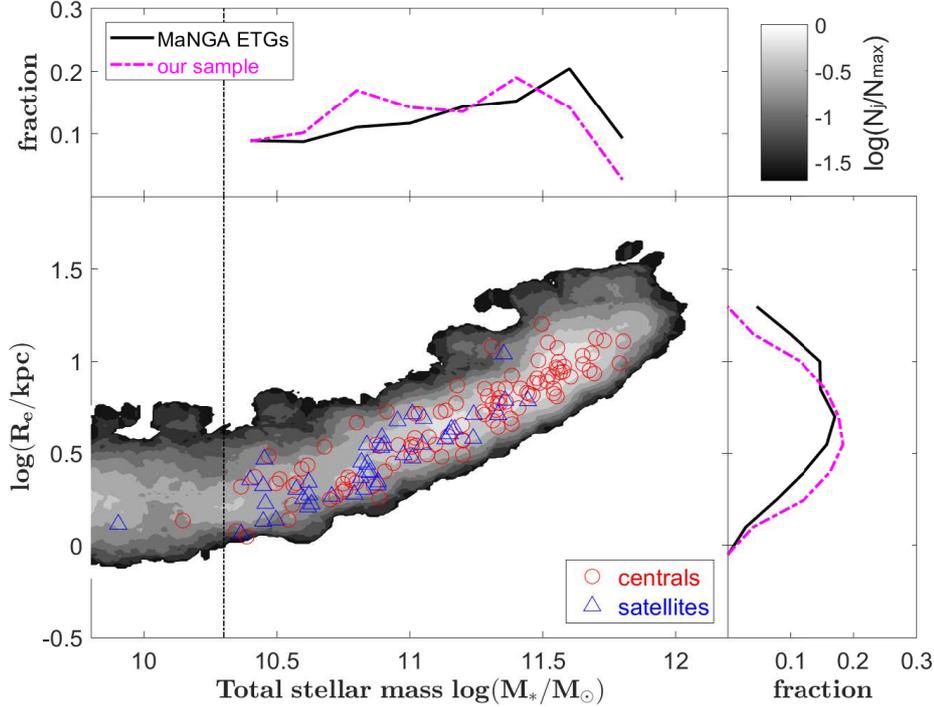}
    \caption{The distribution of our sample on a total stellar mass $M_*$ versus effective radii $R_{\rm e}$ plot. The red circles represent central ETGs and the blue triangles are for satellite ETGs. The vertical black dashed line indicates the stellar mass $\log(M_*/M_{\odot})=10.3$. The background colour map shows the number count distribution of all ETGs in the MaNGA MPL6 with stellar mass $\log(M_*/M_{\odot})>9.7$, with their number densities $\log(N_j/N_{\rm max})$ being indicated by the colour bar. $N_j$ is the number of galaxies in mass vs effective radius bin $j$, and $N_{\rm max}$ is the maximum of the $N_j$. We calculate the marginalised fractions of galaxies with $\log(M_*/M_{\odot})\textgreater10.3$ by mass and size, and show them in the figure. The black solid lines are for overall ETGs while the magenta dashed lines are for our sample.The two samples are slightly different on the marginalized mass and size distribution in the mass range of $10.3\textless\log(M_*/M_{\odot})\textless11.8$, but follow the same mass-size relation.}
    \label{sample}
\end{centering}
\end{figure*}

We select central galaxies and satellites by matching the galaxy catalogue from MaNGA MPL4 (MPL = MaNGA Product Launches; MPL4 contains a total of 1390 galaxies) and the galaxy group catalogue \citep{Yang2007} of SDSS DR7 (Data Release). The brightest galaxy in a group is tagged as the central galaxy and all others as satellite galaxies \citep{Yang2008}. In addition, if the brightest galaxy is not the most massive one in a group, we remove all the galaxies in this group to avoid any possible effects caused by confusing centrals and satellites. In total we have 907 central and 357 satellite galaxies. As the data quality of low luminous galaxies is not good enough for our modelling, we manually choose galaxies with SDSS r-band model magnitude brighter than 15, and arrive at 290 central galaxies and 122 satellites. Using the ETG classification from Galaxy Zoo 1 \citep{Lintott2011}, 141 central and 58 satellite ETGs are then selected. From a visual inspection of these ETGs based on both SDSS images and kinematic maps in MaNGA MPL5 data-analysis pipeline (DAP, \citealp{Westfall2019}), we reduce our candidate galaxies further still. The main criteria are as follows. Firstly, we exclude galaxies that are merging, or have ring structures, dust lanes, foreground stars or anything else that will affect data quality or is hard to model. Secondly, where we have multiple observations of the same galaxy, we choose the observations with the most measurement points. Finally we arrive at 105 central and 44 satellite ETGs and we take these as our sample.

We take the Petrosian half-light radii in the NASA-Sloan Atlas catalog \citep{Blanton2009} as the effective radii $R_{\rm e}$ of our galaxies. We show the distribution of our sample on a population synthesis total stellar mass $M_*$ (see $\S$~\ref{sec4.3} for details) versus effective radius $R_{\rm e}$ plot in Fig.~\ref{sample}, where we also compare our sample with the complete MaNGA ETG sample in MPL6. The two samples are slightly different on the marginalized mass and size distribution in the mass range of $10.3\textless\log(M_*/M_{\odot})\textless11.8$, but follow the same mass-size trend, which means our sample from the early MPL4 catalogue is representative of the later, larger MPL6 ETG sample. The stellar masses of the central ETGs range from $10^{10.15}$ to $10^{11.80}M_{\odot}$, while satellites range from $10^{9.90}$ to $10^{11.44}M_{\odot}$.

\subsection{Individual galaxy data}
\label{sec3.3}
\begin{figure*}
\begin{centering}
	\includegraphics[width=10cm]{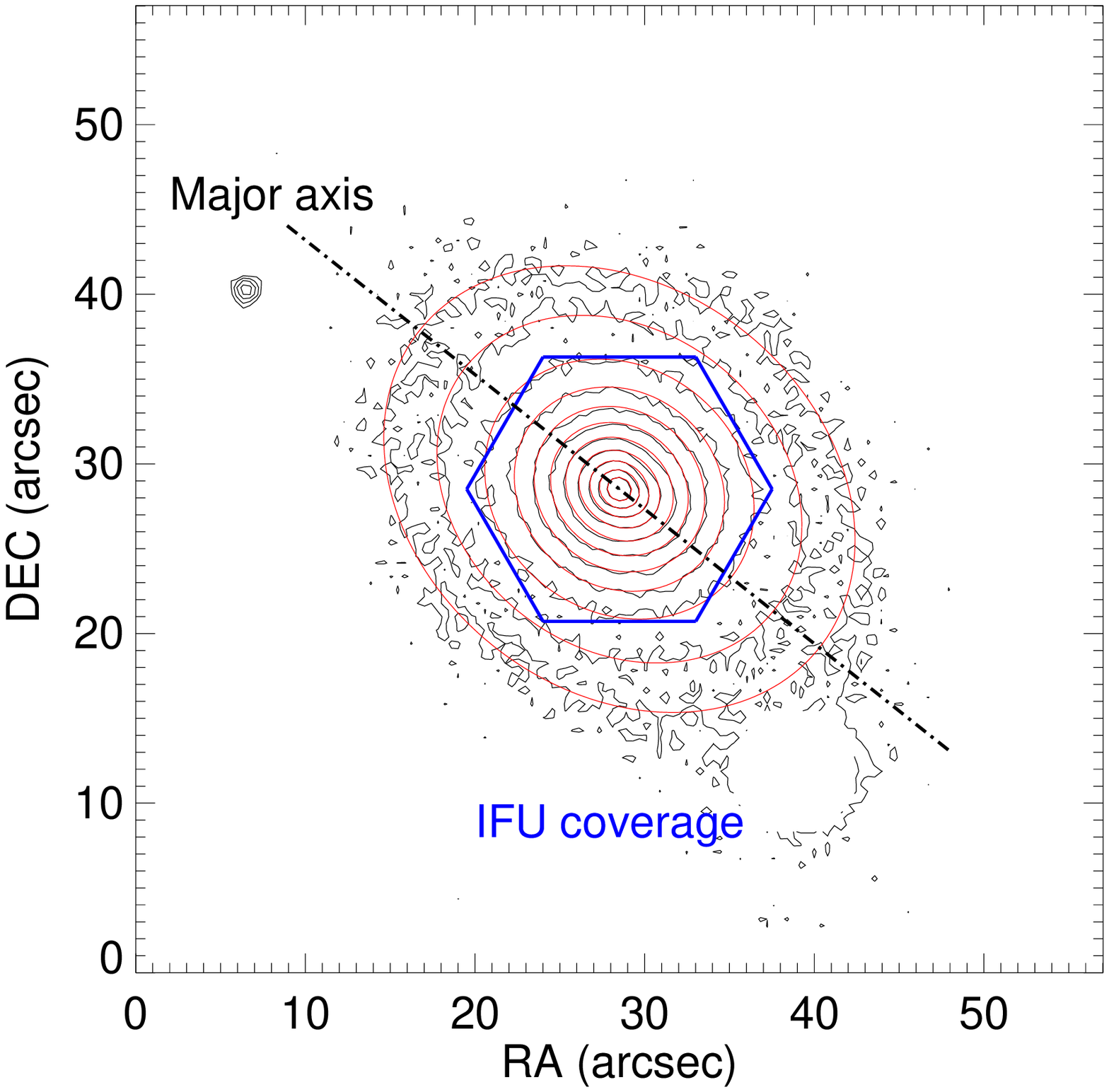}
    \caption{Surface brightness contours and MGE fitting results for example MaNGA galaxy 8247-6101 (plate-IFU). The blue hexagon shows the coverage of the IFU fibres. The black contours represent the original image while the red contours are from the MGE model. The contour interval is equal to 0.5 magnitude. The black dashed line indicates the photometric major axis.}
    \label{img-mge}
\end{centering}
\end{figure*}
\begin{figure}
\begin{centering}
	\includegraphics[width=8.5cm]{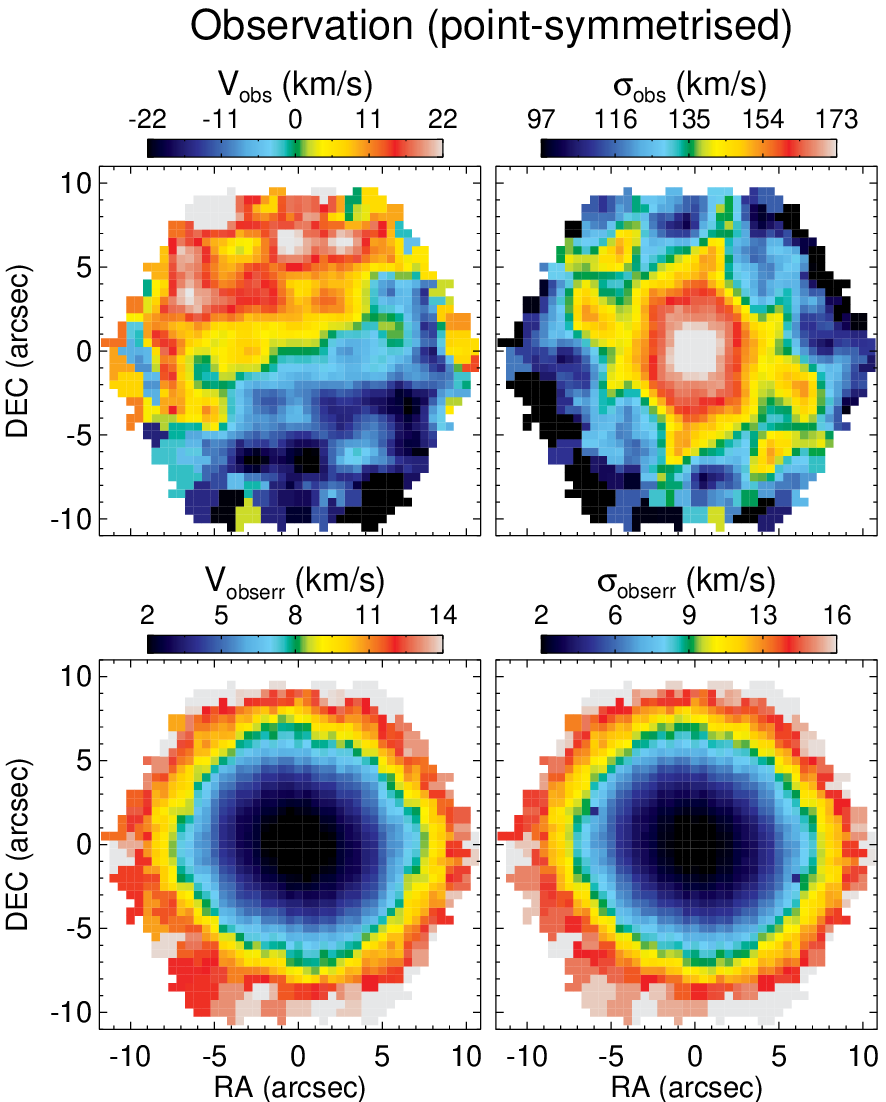}
    \caption{The point-symmetrised observational kinematic data for galaxy 8247-6101. The first row shows the line of sight mean velocity $V_{\rm obs}$ and velocity dispersion $\sigma_{\rm obs}$, while the second row shows the corresponding observational errors, $V_{\rm obserr}$ and $\sigma_{\rm obserr}$.}
    \label{kin-data-obs}
\end{centering}
\end{figure}

As mentioned in \citet{Jin2019}, the Schwarzschild implementation uses both luminosity and kinematics as model constraints, including the line of sight mean velocity $V$ and velocity dispersion $\sigma$. For the luminosity constraints, we apply the MGE method \citep{Cappellari2002} to model the SDSS r-band image mosaics. The MGE fitted two-dimensional surface brightness together with the deprojected three-dimensional luminosity density are used as our model luminosity constraints. For the kinematic constraints, we use data from MaNGA MPL5 (internal release of SDSS DR14, \citealp{Abolfathi2018}) which was generated by the MaNGA data-analysis pipeline (DAP, \citealp{Westfall2019}). We show our model constraints for an example MaNGA galaxy, 8247-6101 (plate-IFU), in Fig.~\ref{img-mge} for the surface brightness and contours from the MGE fitting process, and in Fig.~\ref{kin-data-obs} for the point-symmetrised mean velocity $V$ and velocity dispersion $\sigma$.

\section{Modelling}
\label{sec4}
\begin{figure*}
\begin{centering}
	\includegraphics[width=16cm]{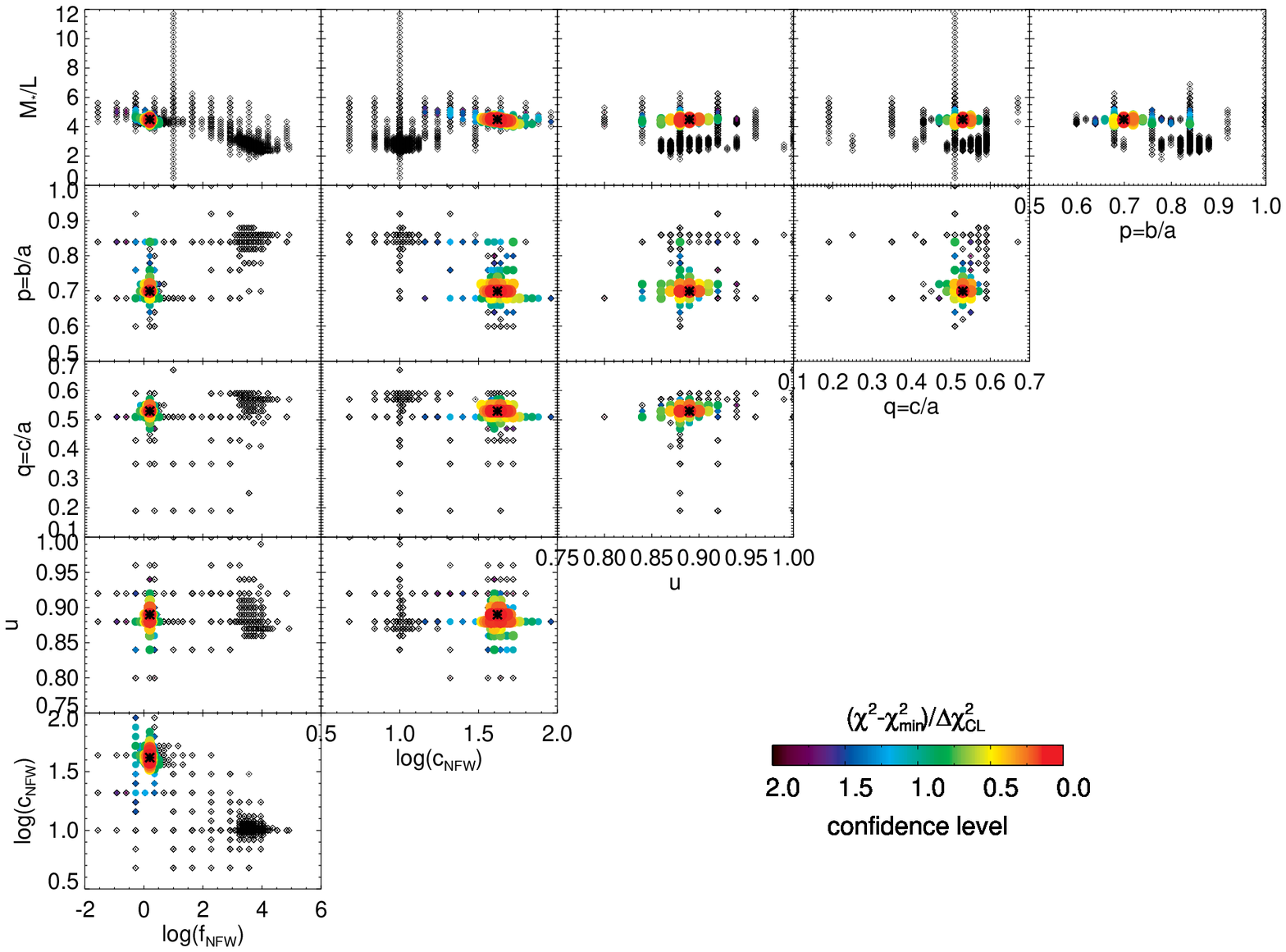}
    \caption{The parameter space of a complete set of modelling runs for galaxy 8247-6101. There are in total six free parameters: the stellar mass to light ratio $M_*/L$, the medium to major axis ratio $p=b/a$, the minor to major axis ratio $q=c/a$, the compression factor $u$, the dark matter concentration $c_{\rm NFW}$ and the virial mass in units of the total stellar mass $f_{\rm NFW}=M_{\rm 200}/M_*$. The dots represent the parameters we have explored. The largest red dot with an asterisk inside indicates the best-fitting model and the other coloured dots indicate models within the $2\sigma$ confidence level, as indicated by the colour bar. The small black dots represent models outside of the $2\sigma$ confidence level (see $\S$~\ref{sec4.2} for the definition of confidence level).}
    \label{grids}
\end{centering}
\end{figure*}
\begin{figure}
\begin{centering}
	\includegraphics[width=8.5cm]{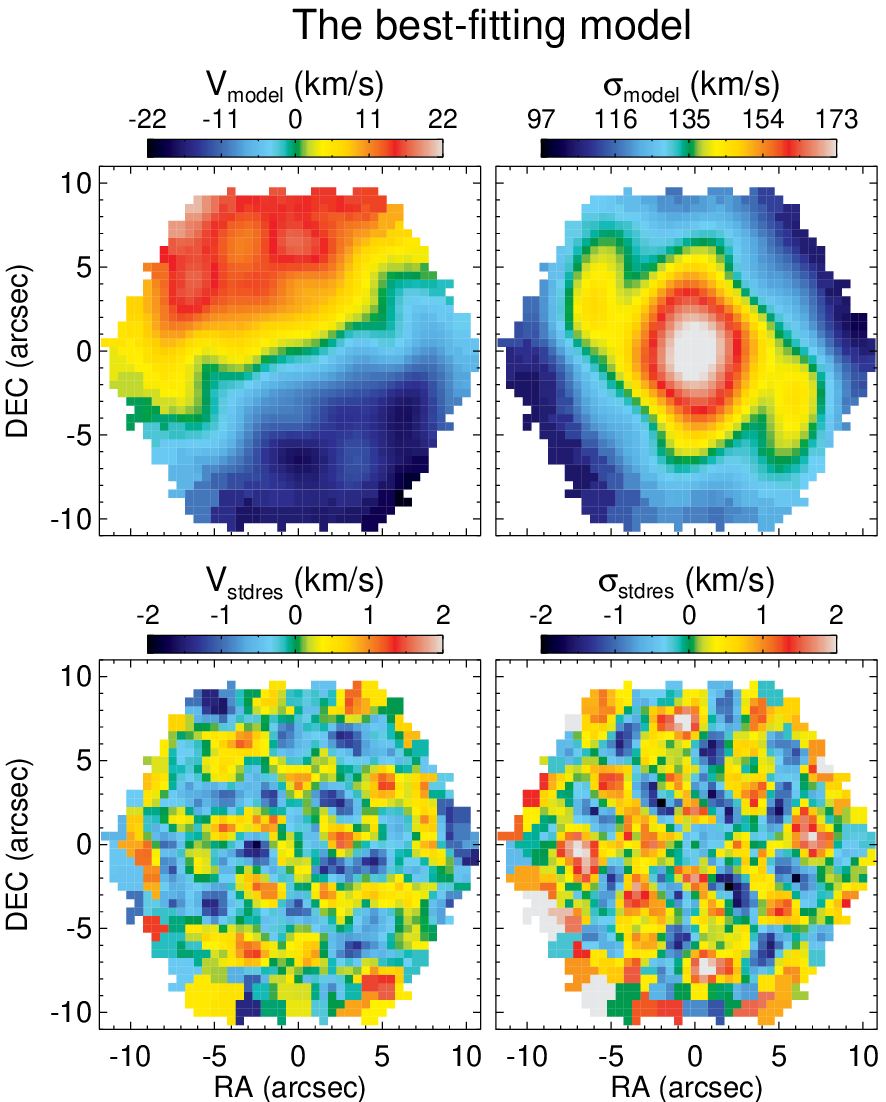}
    \caption{Kinematic data from the best-fitting model for galaxy 8247-6101 and related residual maps. The first row shows the mean velocity $V_{\rm model}$ and velocity dispersion $\sigma_{\rm model}$ maps, while the second row shows the corresponding standardized residual maps between the model and observational data: $V_{\rm stdres}=(V_{\rm model}-V_{\rm obs})/V_{\rm obserr}$ and $\sigma_{\rm stdres}=(\sigma_{\rm model}-\sigma_{\rm obs})/\sigma_{\rm obserr}$.}
    \label{kin-data-model}
\end{centering}
\end{figure}

As indicated earlier, we are using two modelling regimes, Schwarzschild modelling and stellar population synthesis.
The Schwarzschild modelling method that we
use in this work is fully described in \citet{Jin2019}. Here, we describe how the values of free parameters and confidence levels are found using Schwarzschild's method. The grid search of free parameters and information about the best-fitting model are set out in $\S$~\ref{sec4.1}. The confidence levels we describe in $\S$~\ref{sec4.2} are used for estimating statistical errors in our model parameters.

We use stellar population synthesis to calculate the galaxy stellar masses we use for comparison purposes.
In $\S$~\ref{sec4.3}, we present how we calculate these stellar masses.

\subsection{Parameter grids and the best-fitting model}
\label{sec4.1}
We have a total of six free parameters in our models. Three are intrinsic stellar shape parameters: the medium to long axis ratio $p=b/a$, the short to long axis ratio $q=c/a$, and the major axis compression factor $u=a_{\rm 2D}/a_{\rm 3D}$ between the two-dimensional and three-dimensional Gaussians in the MGE procedure. Two are NFW dark matter halo parameters: the dark matter concentration $c_{\rm NFW}$ and the virial mass in units of the total stellar mass $f_{\rm NFW}=M_{\rm 200}/M_*$. Lastly, we have the stellar mass to light ratio $M_*/L$. Section 4 in \citet{Jin2019} contains full descriptions of these parameters.

We find the best-fitting model by grid searching across the parameter space. The best-fitting model is defined as the model with minimum kinematic $\chi^2$:
\begin{equation}
    \chi^2=\sum_{\rm n=1}^{N_{\rm kin}}\left[ \left(\frac{V^{\rm n}_{\rm model}-V^{\rm n}_{\rm obs}}{V^{\rm n}_{\rm obserr}}\right)^2+\left(\frac{\sigma^{\rm n}_{\rm model}-\sigma^{\rm n}_{\rm obs}}{\sigma^{\rm n}_{\rm obserr}} \right)^2\right],
\label{chi2kin}
\end{equation}
where $V^{\rm n}_{\rm model}$, $\sigma^{\rm n}_{\rm model}$ are the model predictions for the $n$-th bin and $V^{\rm n}_{\rm obs}$, $\sigma^{\rm n}_{\rm obs}$ are the observed values, while $V^{\rm n}_{\rm obserr}$, $\sigma^{\rm n}_{\rm obserr}$ represent the observational errors. $N_{\rm kin}$ is the number of bins in the kinematic maps. We assume several groups of initial parameters as starting points for the Schwarzschild models. After these first models have been run, we adopt an iterative process for exploring the parameter space. We create new models around the existing models with lower kinematic $\chi^2$ values by changing the values of parameters in fixed step sizes. This iteration is repeated until we obtain a $\chi^2$ minimum region. We then reduce the step sizes by half to create more models around this region, and continue the iteration. By restricting the minimum step sizes for the free parameters, we eventually find the best-fitting model from a minimum in the kinematic $\chi^2$, and surrounded by enough models for the evaluation of statistical errors. The parameter space for a complete model run for galaxy 8247-6101 is shown in Fig.~\ref{grids}. The dots represent the parameters we have explored. The largest red dot with an asterisk inside indicates the best-fitting model, and whose kinematic map is presented in Fig.~\ref{kin-data-model}. Other coloured and small black dots indicate respectively the models within and outside the $2\sigma$ confidence level (see $\S$~\ref{sec4.2} for the definition of the confidence level). Note that the axis ratios $p=b/a$ and $q=c/a$ here correspond to the three-dimensional deprojection of the flattest Gaussian, and not the average axis ratios of this galaxy.

\subsection{Confidence levels}
\label{sec4.2}
In the Schwarzschild method, using the standard definition of $\Delta\chi^2$ as a confidence level is not feasible, because the models are non-linear in the model parameters, and, from a practical perspective, model fluctuations dominate the variation in $\chi^2$. We choose to follow our earlier paper \citet{Jin2019}, which is based on \citet{RvdB2008}, but with an added re-scale factor of 2 before the square root term, and define the $1\sigma$ confidence level as
\begin{equation}
    \Delta\chi^2 \equiv \chi^2-\chi^2_{\rm min}<2\times\sqrt{2(N_{\rm obs}-N_{\rm par})} \equiv \Delta\chi^2_{\rm CL},
\label{CL}
\end{equation}
where $\chi^2_{\rm min}$ means the chi-square of the best-fitting model. $N_{\rm obs}$ is the number of kinematic constraints ($N_{\rm obs}=2N_{\rm kin}$ as we use $V$ and $\sigma$ as model constraints). $N_{\rm par}$ is the number of free parameters ($N_{\rm par}=6$ here). The models whose $\chi^2$ satisfy this inequality are included for calculating the statistical uncertainties of the model parameters for single data analysis. The maximum and minimum values of the parameters or properties in these models are treated as upper and lower limits in $1\sigma$ error regions.

\subsection{Stellar masses}
\label{sec4.3}
Utilising the algorithm described in \citet{Li2017}, the stellar masses within $R_{\rm e}$ of all the MaNGA galaxies are obtained through stellar population synthesis (SPS) techniques, using the pPXF full spectrum fitting software \citep{Cappellari2004,Cappellari2017}, the MILES stellar libraries \citep{Vazdekis2010}, and by assuming a Chabrier initial mass function (IMF, \citealp{Chabrier2003}). The total stellar mass $M_*$ of a galaxy is estimated as twice the stellar mass within $R_{\rm e}$. The scatter of stellar masses calculated in this way, by assuming a particular IMF, is typically less than 0.1 dex \citep{Li2017,Ge2018}. Assuming a Salpter IMF \citep{Salpeter1955} leads to a systematical shift of 0.25 dex in stellar mass.

As overviewed in $\S$~\ref{sec2}, we are able to obtain the dynamical stellar mass constrained by our Schwarzschild modelling for the galaxies in our sample. For convenient comparison with the whole MaNGA sample and previous results in the literature, we use stellar masses $M_*$ estimated by SPS throughout the paper unless stated otherwise. For our sample, in Fig.~\ref{mass-comparison}, we show a comparison of stellar masses obtained from these two methods for our sample. Using one instead of the other will not significantly change our results. However, this approach of mixing data from different modelling regimes is not satisfactory and we discuss the matter further in $\S$~\ref{sec8}.

\section{Results for an individual galaxy}
\label{sec5}
\begin{figure}
\begin{centering}
	\includegraphics[width=8.5cm]{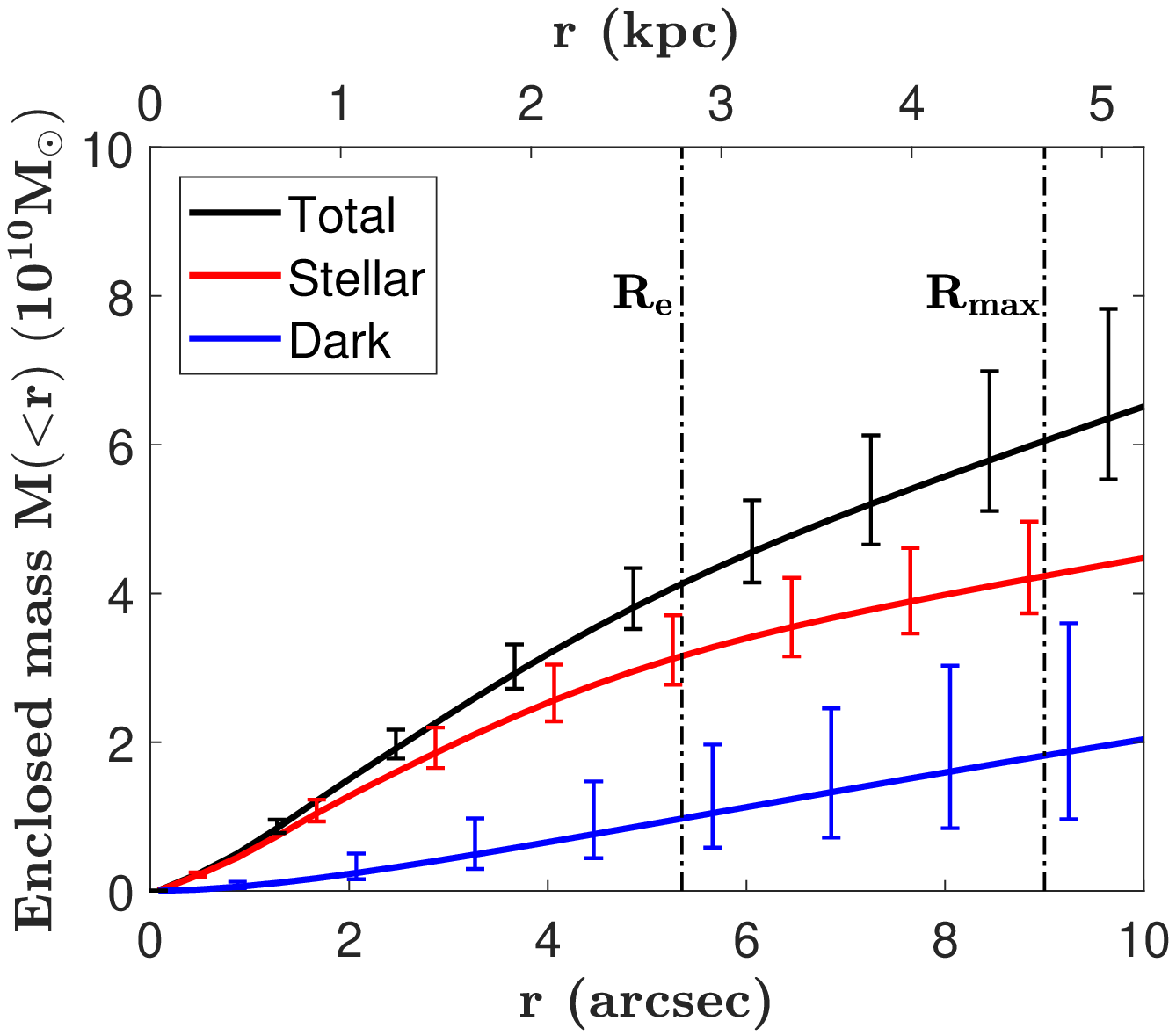}
    \caption{Mass profiles for the example galaxy 8247-6101 inferred from Schwarzschild modelling. The black curve indicates the mass profile, while the red and blue curves represent stellar mass and dark matter respectively. The error bars represent the $1\sigma$ confidence levels. The vertical dashed lines show the effective radius $R_{\rm e}$ and the maximum IFU coverage $R_{\rm max}$.}
    \label{example-mass}
\end{centering}
\end{figure}
\begin{figure}
\begin{centering}
	\includegraphics[width=8.5cm]{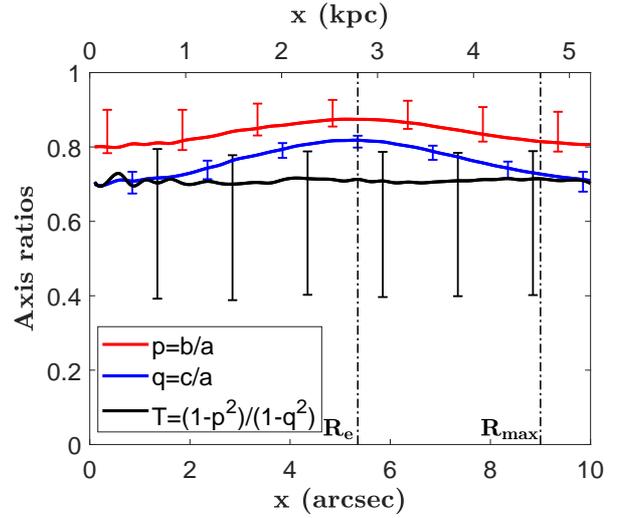}
    \caption{Variation along the major axis ($x$) of the axial ratios $p=b/a, q=c/a$ and triaxial parameter $T=(1-p^2)/(1-q^2)$ for the example galaxy 8247-6101. The red, blue and black curves correspond to $p$, $q$ and $T$. The error bars represent the $1\sigma$ confidence levels. The vertical dashed lines indicate the effective radius $R_{\rm e}$ and the maximum IFU coverage $R_{\rm max}$.}
    \label{example-shape}
\end{centering}
\end{figure}
\begin{figure}
\begin{centering}
	\includegraphics[width=8.5cm]{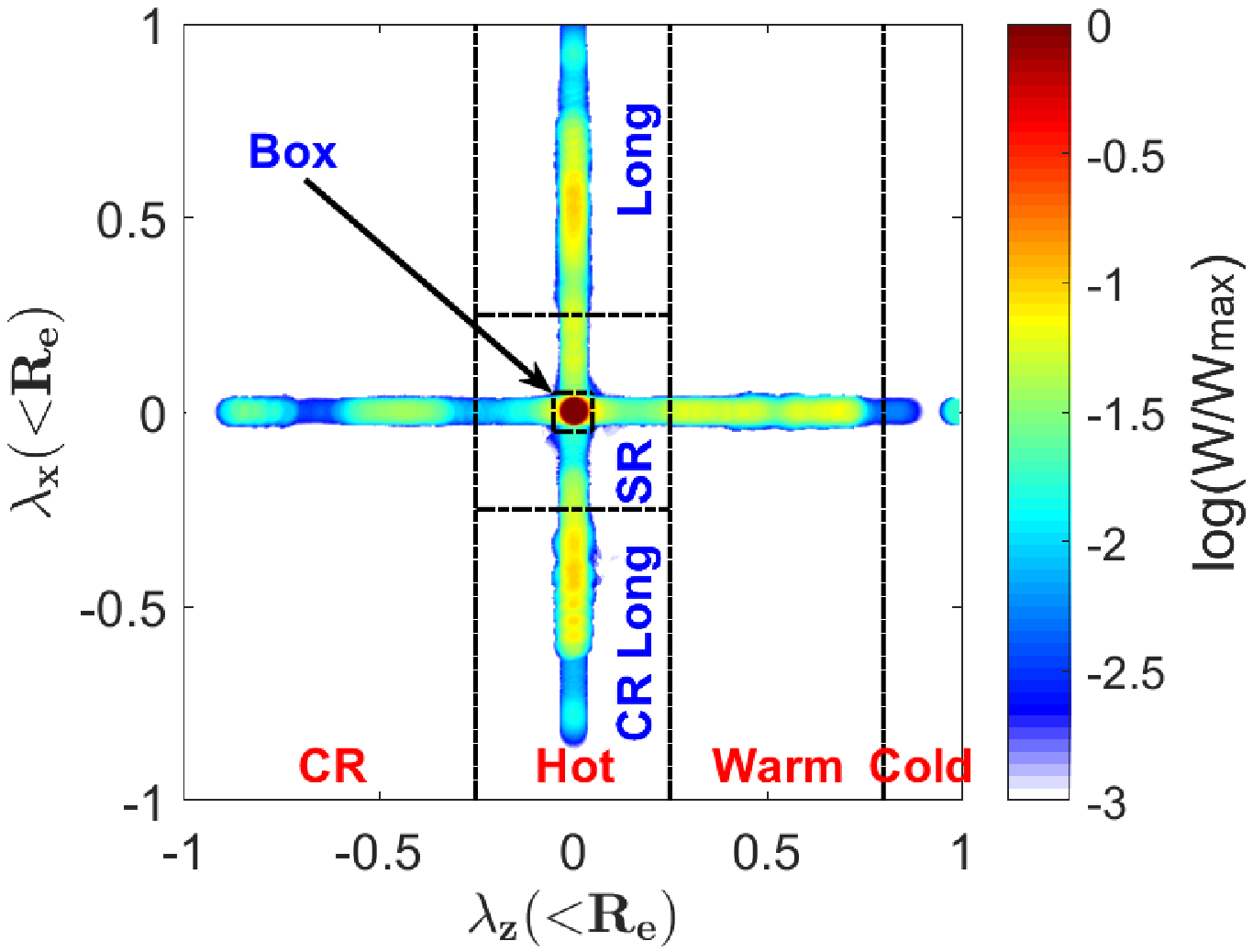}
    \caption{Binned orbit weights by classification within one $R_{\rm e}$ on the $\lambda_z-\lambda_x$ plane for galaxy 8247-6101. The binned values, $\log(W_{\rm j}/W_{\rm max})$, are indicated by the colour bar. $W_{\rm j}$ is the total orbit weight in $\lambda_z$ vs $\lambda_x$ bin $j$, and $W_{\rm max}$ is the maximum of the $W_{\rm j}$. The black dashed lines show the classification of orbits into different categories. Orbits with different $\lambda_z$ values range from cold ($\lambda_z\ge0.8$), warm ($0.25<\lambda_z<0.8$), hot ($-0.25\le\lambda_z\le0.25$) and counter-rotating orbits ($\lambda_z<-0.25$). Hot orbits are separated into four  components: prograde long-axis tubes ($\lambda_x>0.25,\left|\lambda_z\right|\le0.25$), counter-rotating long-axis tubes ($\lambda_x<-0.25,\left|\lambda_z\right|\le0.25$), box orbits ($\left|\lambda_x\right|\le0.05,\left|\lambda_z\right|\le0.05$) and slowly-rotating orbits ($\left|\lambda_x\right|,\left|\lambda_z\right|\le0.25,\left|\lambda_x\right|$ or $\left|\lambda_z\right|>0.05$). Note that the divisions of long-axis tubes and box orbits are just approximations and not the exact.}
    \label{example-circularity}
\end{centering}
\end{figure}

In this section, we show the detailed results for one galaxy. It is not practical to show the detailed results for all 149 galaxies in our sample. Given our objectives in $\S$~\ref{sec1}, we focus on the following galaxy properties: the mass distribution including both stellar and dark matter components, the intrinsic stellar shape, and the internal orbit circularity distribution. In this section, we continue to take galaxy 8247-6101 as an example and show its detailed model results.

Fig.~\ref{example-mass} shows the mass profiles from the best-fitting model for our example galaxy (the black hole mass is negligible). The black, red and blue lines indicate the profiles of the total mass $M_{\rm tot}(\textless r)$, stellar mass $M_*(\textless r)$ and dark matter $M_{\rm dark}(\textless r)$, respectively. The error bars represent the $1\sigma$ confidence level defined as in $\S$~\ref{sec4.2}. The effective radius $R_{\rm e}$ of this galaxy is 5.35 arcsec (2.79 kpc) and the maximum IFU coverage $R_{\rm max}$ (the blue hexagon in Fig.~\ref{img-mge}) is 9 arcsec.

We present the variation of intrinsic stellar shape as a function of radius $r$ for the best-fitting model in Fig.~\ref{example-shape}. The red line represents the medium to long axis ratio $p(r)=b/a$ and the blue line indicates the short to long axis ratio $q(r)=c/a$. The triaxial parameter $T(r)=(1-p(r)^2)/(1-q(r)^2)$ \citep{Binney2008}, which indicates galaxy triaxiality, is shown by the black line. The error bars have the same meaning as Fig.~\ref{example-mass}.

The orbit circularity parameters $\lambda_z$ and $\lambda_x$, which represent the rotations about the minor axis ($z$-axis) and the major axis ($x$-axis), are defined as a ratio of time-averaged quantities \citep{Zhu2018a,Zhu2018b,Jin2019}
\begin{equation}
\left\{
\begin{array}{ll}
    \lambda_z=\overline{L_z}/(\overline{r}\times\overline{V_{\rm rms}}),&\\[1mm]
    \lambda_x=\overline{L_x}/(\overline{r}\times\overline{V_{\rm rms}}),
\end{array}
\right.
\end{equation}
where $\overline{L_z}=\overline{xv_y-yv_x}$, $\overline{L_x}=\overline{yv_z-zv_y}$, $\overline{r}=\overline{\sqrt{x^2+y^2+z^2}}$ and $\overline{V_{\rm rms}}=\sqrt{\overline{v_x^2+v_y^2+v_z^2+2v_xv_y+2v_xv_z+2v_yv_z}}$. Typical short-axis tube orbits have $0\textless\lambda_z\textless 1$ and $\lambda_x\sim0$, while for long-axis tube orbits we have $\lambda_z\sim0$ and $0\textless\lambda_x\textless 1$. Box orbits that dominate by centrophilic orbits satisfy both $\lambda_z\sim0$ and $\lambda_x\sim0$. Negative values of orbit circularity mean the orbits are counter-rotating. We calculate the circularity parameters for each orbit. Based on the orbit weights from the best-fitting model, we obtain the probability density distribution of the orbits with $\overline{r}\textless R_{\rm e}$ on the $\lambda_z-\lambda_x$ plane shown in Fig.~\ref{example-circularity}. Following \citet{Jin2019}, we first divide orbits with different $\lambda_z$ into cold ($\lambda_z\ge0.8$), warm ($0.25<\lambda_z<0.8$), hot ($-0.25\le\lambda_z\le0.25$) and counter-rotating orbits ($\lambda_z<-0.25$). In order to distinguish long-axis tube orbits and box orbits from hot components, we separate hot orbits with a different $\lambda_x$ to be prograde long-axis tubes ($\lambda_x>0.25,\left|\lambda_z\right|\le0.25$), counter-rotating long-axis tubes ($\lambda_x<-0.25,\left|\lambda_z\right|\le0.25$), box orbits ($\left|\lambda_x\right|\le0.05,\left|\lambda_z\right|\le0.05$) and slowly-rotating orbits ($\left|\lambda_x\right|,\left|\lambda_z\right|\le0.25,\left|\lambda_x\right|$ or $\left|\lambda_z\right|>0.05$).

Taking Fig.~\ref{example-shape} into consideration, we find this galaxy is a prolate-like triaxial galaxy ($T\sim0.7$). Examining the orbital separation as shown in Fig.~\ref{example-circularity}, this galaxy has about 14 percent short-axis tube orbits (cold and warm components), 6 percent counter-rotating short-axis tubes, 17 percent long-axis tubes and 16 percent counter-rotating long-axis tubes. Neither rotation about the minor axis nor rotation about the major axis dominate this galaxy. These results are consistent with the model constraints shown in Fig.~\ref{img-mge} and Fig.~\ref{kin-data-obs}. The galaxy rotates slowly ($V\sim20$ km/s compared to $\sigma\sim130$ km/s), and there is a significant misalignment ($\rm \Delta PA=\left|PA_{kin}-PA_{pho}\right|=28^{\circ}$) between the kinematic position angle $\rm PA_{kin}$ and photometric position angle $\rm PA_{pho}$. Such misalignments typically appear in the observation of triaxial galaxies.

\section{Statistical results}
\label{sec6}
In this section, we show the composite statistical results from analysing all the central and satellite ETGs in our sample. We describe how we assess model uncertainties in $\S$~\ref{sec6.1}, mass distributions are presented in $\S$~\ref{sec6.2}, intrinsic stellar shapes in $\S$~\ref{sec6.3}, and orbit circularity distributions in $\S$~\ref{sec6.4}. In these sub-sections, we create the model determined galaxy parameter distributions by binning the parameter values using equal count bins. The parameter distributions, separated into central and satellite ETGs, are shown in Fig.~\ref{DMfrac-vs-mass}, Fig.~\ref{shape-vs-mass} and Fig.~\ref{lambda-z}.

\subsection{Model uncertainties}
\label{sec6.1}
\begin{table*}
\caption{The systematic uncertainties $\sigma_{\rm sys}(X)$ and systematic biases $\mathcal{D}(X)$ for oblate, triaxial and prolate ETGs in \citet{Jin2019}. From left to right, the properties are: (1) the dark matter fraction within one effective radius $f_{\rm DM}(\textless R_{\rm e})$ (see $\S$~\ref{sec6.2}); (2) the triaxial parameter at one effective radius $T_{\rm e}=(1-p_{\rm e}^2)/(1-q_{\rm e}^2)$ (see $\S$~\ref{sec6.3}); (3) the fractions of orbits $f_{\rm cold}$, $f_{\rm warm}$, $f_{\rm hot}$ and $f_{\rm CR}$ (see $\S$~\ref{sec6.4}); (4) the fractions of orbits $f_{\rm prolong}$, $f_{\rm CRlong}$, $f_{\rm box}$ and $f_{\rm SR}$ (see $\S$~\ref{sec6.4}). The systematic biases of dark matter fractions are hard to estimate, as they depend on how close the NFW halo we used is to the real dark matter halo of the galaxy (see \citealp{Jin2019} for details).}
    \centering
	\begin{tabular}{|c|c|c|c|c|c|c|c|c|c|c|c|}
    \hline
    \multicolumn{2}{|c|}{\multirow{2}*{$\sigma_{\rm sys}(X)$}} & \multicolumn{10}{c|}{Property $X$}\\
    \cline{3-12}
    \multicolumn{2}{|c|}{} & $f_{\rm DM}(\textless R_{\rm e})$ & $T_{\rm e}$ & $f_{\rm cold}$ & $f_{\rm warm}$ & $f_{\rm hot}$ & $f_{\rm CR}$ & $f_{\rm prolong}$ & $f_{\rm CRlong}$ & $f_{\rm box}$ & $f_{\rm SR}$\\
    \hline
    \multirow{3}*{Morphology} & oblate & 0.144  & 0.144  & 0.054  & 0.119  & 0.129  & 0.033  & 0.018  & 0.020  & 0.075  & 0.081\\
    ~ & triaxial & 0.115  & 0.267  & 0.034  & 0.116  & 0.161  & 0.064  & 0.078  & 0.067  & 0.104  & 0.082\\
    ~ & prolate & 0.099  & 0.187  & 0.018  & 0.071  & 0.095  & 0.051  & 0.100  & 0.098  & 0.168  & 0.085\\
    \hline
    \hline
    \multicolumn{2}{|c|}{\multirow{2}*{$\mathcal{D}(X)$}} & \multicolumn{10}{c|}{Property $X$}\\
    \cline{3-12}
    \multicolumn{2}{|c|}{} & $f_{\rm DM}(\textless R_{\rm e})$ & $T_{\rm e}$ & $f_{\rm cold}$ & $f_{\rm warm}$ & $f_{\rm hot}$ & $f_{\rm CR}$ & $f_{\rm prolong}$ & $f_{\rm CRlong}$ & $f_{\rm box}$ & $f_{\rm SR}$\\
    \hline
    \multirow{3}*{Morphology} & oblate & / & 0.023  & 0.022  & 0.101  & -0.185  & 0.061  & 0.009  & 0.021  & -0.084  & -0.136\\
    ~ & triaxial & / & -0.003  & -0.006  & 0.078  & -0.129  & 0.056  & 0.036  & 0.031  & -0.089  & -0.116\\
    ~ & prolate & / & -0.014  & 0.014  & 0.043  & -0.100  & 0.044  & 0.008  & 0.012  & 0.016  & -0.149\\
    \hline
    \end{tabular}
\label{uncertainty}
\end{table*}

For an individual galaxy, the model uncertainties in our analysis have three components: a statistical uncertainty $\sigma_{\rm stat}$, a systematic uncertainty $\sigma_{\rm sys}$, and a systematic bias $\mathcal{D}$. The statistical uncertainty $\sigma_{\rm stat}$ is related to the confidence levels which are dominated by the model fluctuations as explained in $\S$~\ref{sec4.2}. For a given galaxy property $X$, we calculate its maximum value $X_{\rm max}$, minimum value $X_{\rm min}$ and best-fitting value $X_{\rm best}$ in the models within a $1\sigma$ confidence level. Thus we define the statistical upper and lower errors as $\sigma_{\rm stat+}(X)=X_{\rm max}-X_{\rm best}$ and $\sigma_{\rm stat-}(X)=X_{\rm best}-X_{\rm min}$.

The systematic uncertainty $\sigma_{\rm sys}$ and systematic bias $\mathcal{D}$ are evaluated based on the model tests against the Illustris simulations \citep{Jin2019}, as first introduced in $\S$~\ref{sec2}. We calculate the residuals of different galaxy properties between the model estimates and the true values for all mock data sets in \citet{Jin2019}. The systematic uncertainty $\sigma_{\rm sys}$ represents the standard deviations of these residuals, while the systematic biases $\mathcal{D}$ are the average values of the residuals. For a given galaxy property $X$, we calculate $\sigma_{\rm sys}(X)$ and $\mathcal{D}(X)$ for oblate galaxies ($T_{\rm model}\le0.3$), triaxial galaxies ($0.3<T_{\rm model}<0.7$) and prolate galaxies ($T_{\rm model}\ge0.7$) separately, where $T_{\rm model}$ represents the triaxial parameter at one $R_{\rm e}$ found by the model. In Table~\ref{uncertainty}, we list the values of $\sigma_{\rm sys}(X)$ and $\mathcal{D}(X)$, with the exception of the systematic biases of the dark matter fractions as these depend on how close the NFW halo we used is to the real dark matter halo of a galaxy and are difficult to estimate (see \citealp{Jin2019} for details).

For the average values of property $X$ binned by $N$ galaxies, we calculate their upper and lower overall uncertainties $U^+(X)$ and $U^-(X)$ as
\begin{equation}
    U^+(X)=U^+_{\sigma}(X)+U^+_{\mathcal{D}}(X),
\label{err1}
\end{equation}
\begin{equation}
    U^-(X)=U^-_{\sigma}(X)+U^-_{\mathcal{D}}(X),
\end{equation}
where the terms with subscript $\sigma$ represent the effect of the statistical uncertainty $\sigma_{\rm stat}(X)$ and the systematic uncertainty $\sigma_{\rm sys}(X)$ to the overall model uncertainties, while the terms with subscript $\mathcal{D}$ indicate the effect of the systematic bias $\mathcal{D}(X)$.

For the uncertainties caused by $\sigma_{\rm stat}(X)$ and $\sigma_{\rm sys}(X)$ in the mean value of property $X$ in a bin of $N$ galaxies, we have
\begin{equation}
    U^+_{\sigma}(X)=\frac{1}{N}\sqrt{\sum_{\rm n=1}^{N}\left[\sigma^2_{\rm stat+}(X,n)+\sigma^2_{\rm sys}(X,n)\right]},
\end{equation}
\begin{equation}
    U^-_{\sigma}(X)=\frac{1}{N}\sqrt{\sum_{\rm n=1}^{N}\left[\sigma^2_{\rm stat-}(X,n)+\sigma^2_{\rm sys}(X,n)\right]},
\end{equation}
where $n$ represents the $n$-th galaxy in the bin.

We consider the effect of systematic bias $\mathcal{D}(X)$ according to its sign. For a bin of $N$ galaxies, we have
\begin{equation}
\left\{
\begin{array}{ll}
    U^+_{\mathcal{D}}(X)=-\overline{\mathcal{D}}(X,n)&\\[1mm]
    U^-_{\mathcal{D}}(X)=0
\end{array}
\right.
if\ \overline{\mathcal{D}}(X,n)<0,
\end{equation}
and
\begin{equation}
\left\{
\begin{array}{lr}
    U^+_{\mathcal{D}}(X)=0&\\[1mm]
    U^-_{\mathcal{D}}(X)=\overline{\mathcal{D}}(X,n)
\end{array}
\right.
if\ \overline{\mathcal{D}}(X,n)>0,
\end{equation}
with
\begin{equation}
    \overline{\mathcal{D}}(X,n)=\frac{1}{N}\sum_{\rm n=1}^{N}\mathcal{D}(X,n).
\label{err2}
\end{equation}
This means that any underestimations in the simulation tests result in a larger upper error bar, as we expect the real values of the property to be larger than our model predictions, while any overestimations, correspondingly, cause a larger lower error bar. The mock galaxies we used in \citep{Jin2019} are generally representative of real observations, so we expect that the method works similarly with the two samples and has an equivalent systematic bias. However, it is difficult to match our real galaxies to mock galaxies one-to-one, and we therefore use the arithmetic mean of biases in this work.

The overall uncertainties $U^+(X)$ and $U^-(X)$ are treated as error bars in the figures plotted in $\S$~\ref{sec6.2}, $\S$~\ref{sec6.3} and $\S$~\ref{sec6.4}.

\subsection{Mass distributions}
\label{sec6.2}
\begin{figure}
\begin{centering}
	\includegraphics[width=8.5cm]{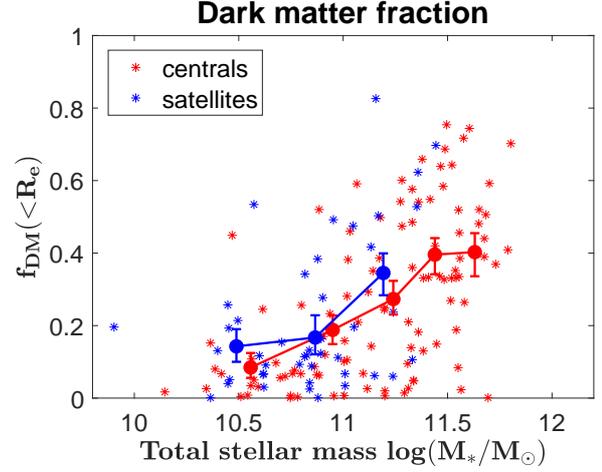}
    \caption{The dark matter fraction within one $R_{\rm e}$ as a function of the galaxies' total stellar mass $M_*$. The red asterisks represent central ETGs while the blue asterisks, satellite ETGs. The red and blue solid lines with error bars show the mean binned values and their uncertainties calculated based on $\S$~\ref{sec6.1}. Here we only consider the statistical errors $\sigma_{\rm stat}$ and systematic errors $\sigma_{\rm sys}$ since the systematic bias $\mathcal{D}$ is hard to evaluate as the true dark matter density slope is not known (see \citealp{Jin2019} for details).}
    \label{DMfrac-vs-mass}
\end{centering}
\end{figure}

The total mass distributions $M_{\rm tot}$ in our investigations include a stellar component $M^{\rm Schw}_*$ and a dark matter component $M_{\rm dark}$ (the mass of a black hole is very small by comparison). We calculate the dark matter fraction within one effective radius $f_{\rm DM}(\textless R_{\rm e})=M_{\rm dark}/M_{\rm tot}$. Fig.~\ref{DMfrac-vs-mass} shows the relationship between the dark matter fraction $f_{\rm DM}(\textless R_{\rm e})$ and the population synthesis total stellar mass $\log(M_*/M_{\odot})$. The red asterisks represent central ETGs while the blue asterisks represent satellite ETGs. The coloured solid lines show the mean binned values and their uncertainties. We can clearly see the dark matter fraction increases with the stellar mass and there is no difference between central and satellite ETGs in the same mass range. Most galaxies around $10^{10.5}M_{\odot}$ have $f_{\rm DM}(\textless R_{\rm e})\textless0.2$, while a lot of massive galaxies with $11.0\textless\log(M_*/M_{\odot})\textless11.5$ have $f_{\rm DM}(\textless R_{\rm e})\textgreater0.4$. This trend is generally consistent with \citet{Cappellari2013a} who used Jeans Anisotropic Modelling (JAM) for $\rm ATLAS^{3D}$ ETGs and found $f_{\rm DM}(\textless R_{\rm e})$ increases with stellar mass when $\log(M_*/M_{\odot})\textgreater10.6$. However, their dark matter fractions are lower than our results for massive galaxies with $f_{\rm DM}(\textless R_{\rm e})\sim0.3$ when using fixed NFW halo based on \citet{Moster2010} and $f_{\rm DM}(\textless R_{\rm e})\sim0.2$ when using free NFW halo. The most massive ETGs are mainly slowly rotating triaxial or prolate galaxies, which do not match the assumptions within JAM modelling very well. In addition, the true dark matter profiles may deviate from NFW profiles. We discuss possible biases for dark matter fractions due to different model assumptions in $\S$~\ref{sec8}.

\subsection{Intrinsic stellar shapes}
\label{sec6.3}
\begin{figure}
\begin{centering}
	\includegraphics[width=8.5cm]{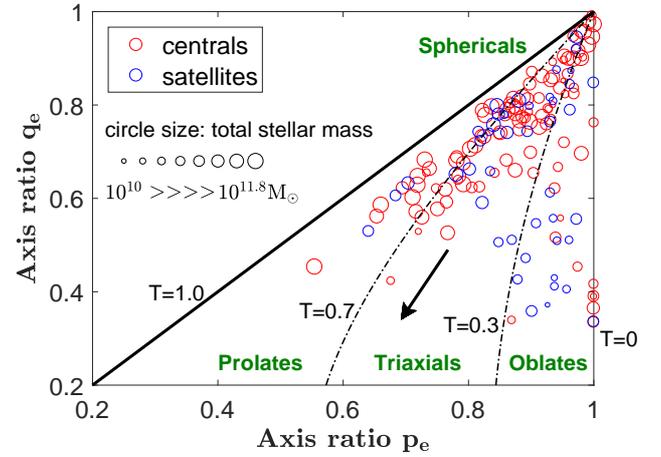}
    \caption{The distribution of axis ratios $p_{\rm e}$ versus $q_{\rm e}$ at one $R_{\rm e}$ from the best-fitting models for all galaxies in our sample. The red circles represent central ETGs while the blue circles represent satellite ETGs. The black dashed curves from right to left indicate $T_{\rm e}=0.3$, $T_{\rm e}=0.7$ and the black thick line is for $T_{\rm e}=1.0$. Larger circle sizes indicate higher total stellar mass. The black arrow indicates the average overestimations $\Delta p_{\rm e}=0.07$ and $\Delta q_{\rm e}=0.14$ according to tests using the Illustris simulations \citep{Jin2019}.}
    \label{shape-2Dmap}
\end{centering}
\end{figure}
\begin{figure}
\begin{centering}
	\includegraphics[width=8.5cm]{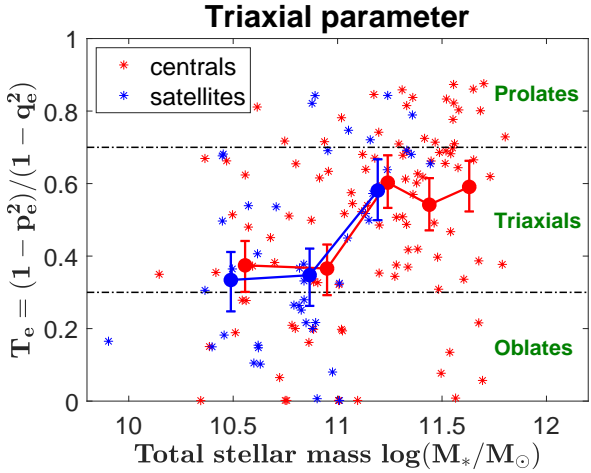}
    \caption{The variation of the triaxial parameter $T_{\rm e}=(1-p_{\rm e}^2)/(1-q_{\rm e}^2)$ as a function of total stellar mass. The red asterisks represent central ETGs while the blue asterisks represent satellite ETGs. The coloured solid lines show the mean binned values and their uncertainties, including the effect of statistical uncertainties, systematic uncertainties and systematic biases as in equations (\ref{err1}) to (\ref{err2}). The horizontal dashed lines indicate $T_{\rm e}=0.3$ and $T_{\rm e}=0.7$.}
    \label{shape-vs-mass}
\end{centering}
\end{figure}

The stellar component in our modelling is taken to be triaxial. As described earlier and in \citet{Jin2019}, we use three parameters to indicate the intrinsic stellar shapes, the medium to long axis ratio $p=b/a$, the short to long axis ratio $q=c/a$ and the major axis compression factor $u=a_{\rm 2D}/a_{\rm 3D}(u\le 1)$, where $a_{\rm 3D}$ is the typical scale of a three-dimensional Gaussian (the standard deviation along major axis) and $a_{\rm 2D}$ is the corresponding typical scale of projected two-dimensional Gaussian. Here we concentrate on $p$ and $q$ as they are directly related to galaxy morphologies. For each galaxy, we obtain the average axis ratio and triaxial parameter profiles as shown in Fig.~\ref{example-shape}. We take $p_{\rm e}$, $q_{\rm e}$ and $T_{\rm e}=(1-p_{\rm e}^2)/(1-q_{\rm e}^2)$ at one effective radius for statistical analysis.

Fig.~\ref{shape-2Dmap} shows the distribution of axis ratios $p_{\rm e}$ versus $q_{\rm e}$ for all galaxies. The red circles represent central ETGs while the blue circles represent satellite ETGs, and the larger symbol sizes indicate larger total stellar masses for the galaxies. The black dashed curves separate the galaxies into oblates ($T_{\rm e}\le0.3$), triaxials ($0.3<T_{\rm e}<0.7$) and prolates ($T_{\rm e}\ge0.7$). The black arrow indicates the average overestimation of $p$ and $q$ in our simulation tests \citep{Jin2019}. We suggest that there should not be so many galaxies that are close to spherical ($p\sim1$ and $q\sim1$). We see that large circles (high stellar masses) tend to located in the prolate regions while the small circles (low stellar masses) tend to located in the oblate regions. In order to check the variation of  intrinsic stellar shapes as a function of stellar mass quantitatively, we show the triaxial parameter $T_{\rm e}$ as a function of total stellar mass $M_*$ in Fig.~\ref{shape-vs-mass}. The red asterisks represent central ETGs while the blue asterisks represent satellite ETGs. The coloured solid lines show the mean binned values and their uncertainties including the effects of statistical uncertainties, systematic uncertainties and systematic biases as in equations (\ref{err1}) to (\ref{err2}). Lower mass galaxies are more oblate-like with their average triaxial parameter $T_{\rm e}\sim0.4$, while higher mass galaxies are more prolate-like with $T_{\rm e}\sim0.6$. There is a rapid change around $\log(M_*/M_{\odot})=11.1$. Both central and satellite ETGs show similar variations of $T_{\rm e}$ as a function of stellar mass $M_*$. The triaxial parameters of central and satellite ETGs are consistent with each other within the $1\sigma$ error bar at the same stellar mass.

\subsection{Orbit circularity distributions}
\label{sec6.4}
\begin{figure*}
\begin{centering}
	\includegraphics[width=16cm]{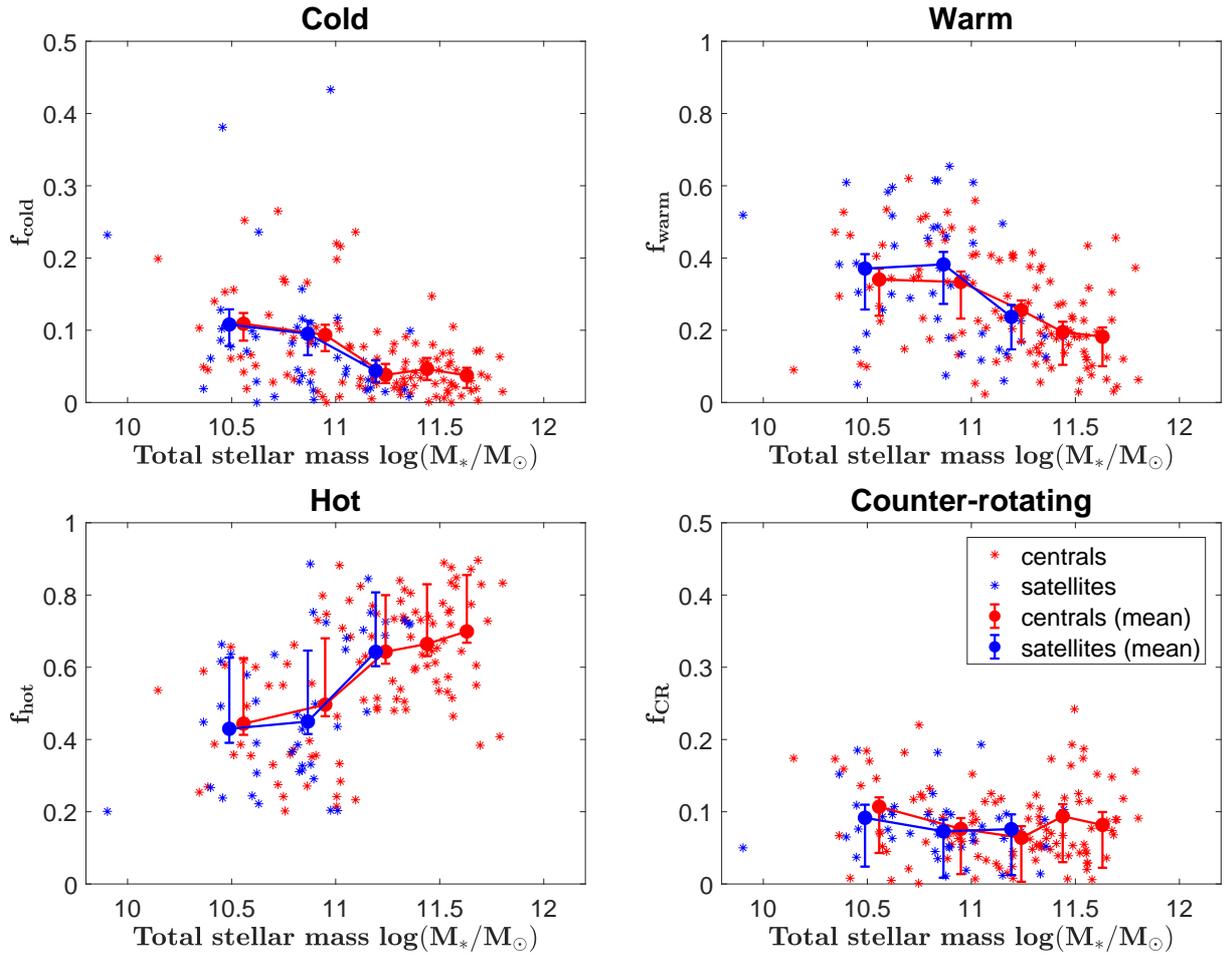}
    \caption{The luminosity fractions of cold, warm, hot and counter-rotating orbits versus total stellar mass. From top-left to bottom-right, they are: cold ($\lambda_z\ge0.8$), warm($0.25<\lambda_z<0.8$), hot($-0.25\le\lambda_z\le0.25$) and counter-rotating ($\lambda_z<-0.25$) components. The red asterisks represent central ETGs while the blue asterisks, satellite ETGs. The coloured solid lines show the mean binned values and their uncertainties, including the effects of statistical uncertainties, systematic uncertainties and systematic biases as in equations (\ref{err1}) to (\ref{err2}).}
    \label{lambda-z}
\end{centering}
\end{figure*}

\begin{figure*}
\begin{centering}
	\includegraphics[width=16cm]{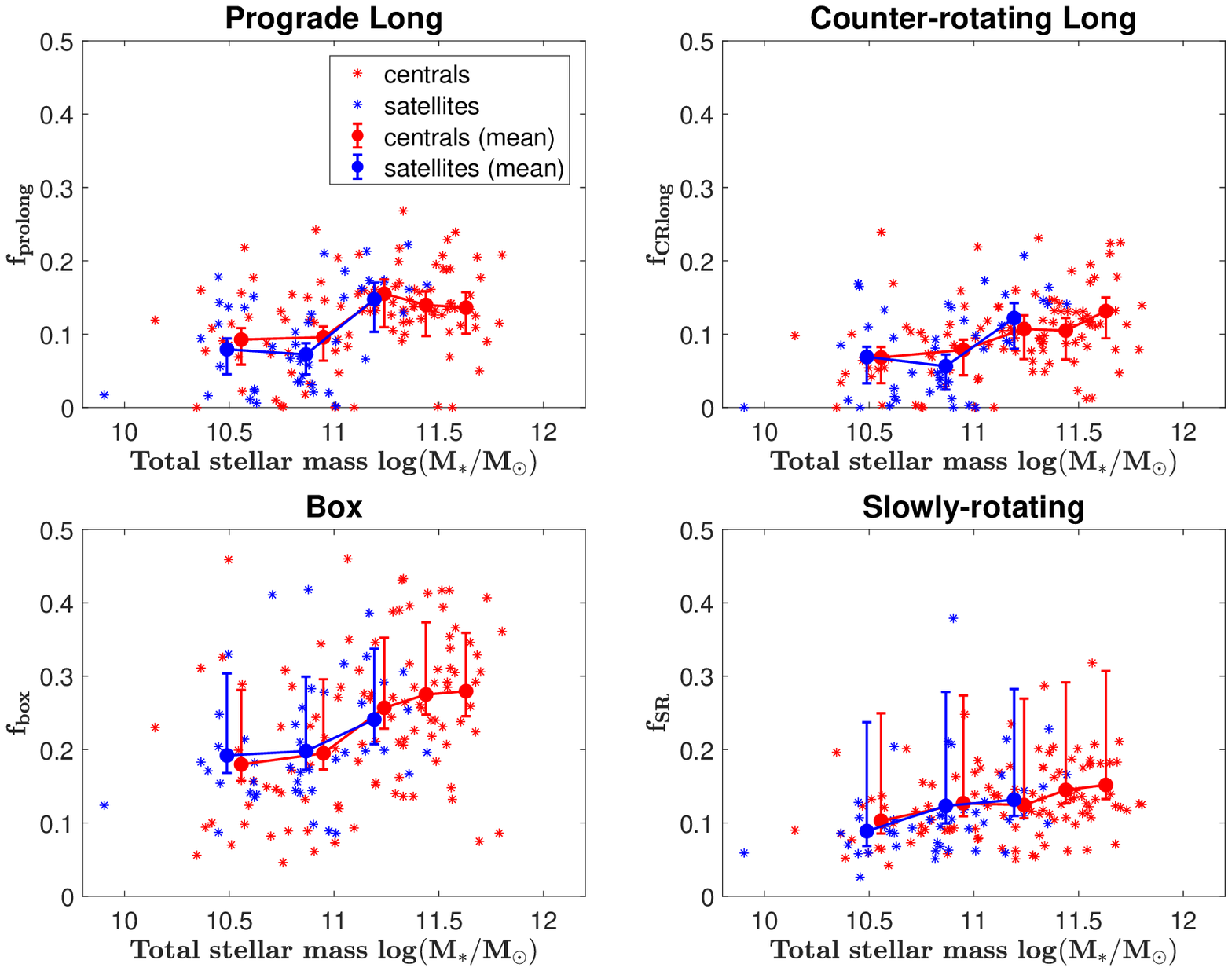}
    \caption{The luminosity fractions of long-axis tubes, box orbits and slowly-rotating orbits versus total stellar mass.
    From top-left to bottom-right, they are: prograde long-axis tube ($\lambda_x>0.25,\left|\lambda_z\right|\le0.25$), counter-rotating long-axis tube ($\lambda_x<-0.25,\left|\lambda_z\right|\le0.25$), box ($\left|\lambda_x\right|\le0.05,\left|\lambda_z\right|\le0.05$) and slowly-rotating ($\left|\lambda_x\right|,\left|\lambda_z\right|\le0.25,\left|\lambda_x\right|$ or $\left|\lambda_z\right|>0.05$) components. The symbols and lines have the same meaning as those in Fig.~\ref{lambda-z}. Note that the divisions of long-axis tubes and box orbits are just approximations and not the exact.}
    \label{lambda-x}
\end{centering}
\end{figure*}

We use the orbit circularity distributions $f(\lambda_z)$ and $f(\lambda_x)$ to illustrate what the internal orbital structures of our galaxies might be like. As shown in Fig.~\ref{example-circularity}, for each galaxy, we obtain the probability density of orbits within one $R_{\rm e}$ and separate them into different components on the $\lambda_z-\lambda_x$ plane. We calculate the luminosity weighted fraction of each component within one $R_{\rm e}$.

Fig.~\ref{lambda-z} shows the fractions of cold ($f_{\rm cold}$, $\lambda_z\ge0.8$), warm ($f_{\rm warm}$, $0.25<\lambda_z<0.8$), hot ($f_{\rm hot}$, $-0.25\le\lambda_z\le0.25$) and counter-rotating orbits ($f_{\rm CR}$, $\lambda_z<-0.25$) versus the total stellar mass. The symbols have the same meaning as Fig.~\ref{shape-vs-mass}. When the stellar mass increases, $f_{\rm cold}$ and $f_{\rm warm}$ decrease while $f_{\rm hot}$ increases. We find that these three fractions all have rapid changes around $\log(M_*/M_{\odot})=11.1$. Galaxies with $\log(M_*/M_{\odot})\textless11.1$ have on average $f_{\rm cold}\sim0.1$, $f_{\rm warm}\sim0.35$ and $f_{\rm hot}\sim0.45$, while galaxies with $\log(M_*/M_{\odot})\textgreater11.1$ have on average $f_{\rm cold}\sim0.05$, $f_{\rm warm}\sim0.2$ and $f_{\rm hot}\sim0.65$. The counter-rotating component generally contributes less than a fraction of $0.1$ and varies little as a function of $M_*$. The variations of $f_{\rm cold}$, $f_{\rm warm}$, $f_{\rm hot}$ and $f_{\rm CR}$ generally have similar trend as the CALIFA ETGs \citep{Zhu2018b}. However, there are significant differences due to the different selection functions of the two samples. We show the comparison of our results and CALIFA results in the appendix.

Following \citet{Jin2019}, we divide the hot component into prograde long-axis tubes ($f_{\rm prolong}$, $\lambda_x>0.25,\left|\lambda_z\right|\le0.25$), counter-rotating long-axis tubes ($f_{\rm CRlong}$, $\lambda_x<-0.25,\left|\lambda_z\right|\le0.25$), box orbits ($f_{\rm box}$, $\left|\lambda_x\right|\le0.05,\left|\lambda_z\right|\le0.05$) and slowly-rotating orbits ($f_{\rm SR}$, $\left|\lambda_x\right|,\left|\lambda_z\right|\le0.25,\left|\lambda_x\right|$ or $\left|\lambda_z\right|>0.05$). The variations of these four components as a function of total stellar mass are shown in Fig.~\ref{lambda-x} and the symbols are the same as Fig.~\ref{shape-vs-mass}. We can see $f_{\rm prolong}$, $f_{\rm CRlong}$, $f_{\rm box}$ and $f_{\rm SR}$ all increase with increasing stellar mass. There is also a rapid change around $\log(M_*/M_{\odot})=11.1$ as revealed in Fig.~\ref{lambda-z}.

Combining Figs.~\ref{lambda-z} and~\ref{lambda-x}, galaxies more massive than $10^{11.1}M_{\odot}$ have a significant fraction of box orbits ($f_{\rm box}\sim0.3$ on average) with centrophilic orbits dominating. However, these galaxies are mainly constructed by tube orbits, with similar amounts of short-axis tube orbits ($f_{\rm cold}+f_{\rm warm}+f_{\rm CR}\sim0.35$) and long-axis tube orbits ($f_{\rm prolong}+f_{\rm CRlong}\sim0.25$). The fraction of counter-rotating orbits ($f_{\rm CR}+f_{\rm CRlong}\sim0.25$) are almost comparable to the co-rotating tube orbits ($f_{\rm cold}+f_{\rm warm}+f_{\rm prolong}\sim0.4$), thus these galaxies have little net rotation. Galaxies less massive than $10^{11.1}M_{\odot}$ are usually dominated by short-axis tube orbits ($f_{\rm cold}+f_{\rm warm}+f_{\rm CR}\sim0.55$), and have less long-axis tube orbits ($f_{\rm prolong}+f_{\rm CRlong}\sim0.15$) and centrophilic box orbits ($f_{\rm box}\sim0.3$), at the same time, counter-rotating orbits ($f_{\rm CR}+f_{\rm CRlong}\sim0.15$) are not remarkable compared to the co-rotating tube orbits ($f_{\rm cold}+f_{\rm warm}+f_{\rm prolong}\sim0.55$), thus these galaxies usually show significant net rotation around the short axis. Not forgetting that orbit distributions are at best only indicative, due to the complicated select function of the MaNGA sample, the orbit distributions of ETGs we obtained here are not generally representative quantitatively of the ETGs in the nearby universe. In both figures, there are no significant differences between central and satellite ETGs regardless of the total stellar mass involved. Since the orbit structures are directly related with galaxy morphologies, it is not surprising that we find similar trends in $\S$~\ref{sec6.3} and $\S$~\ref{sec6.4}.

\section{The effect of local environment}
\label{sec7}
\begin{figure*}
\begin{centering}
	\includegraphics[width=16cm]{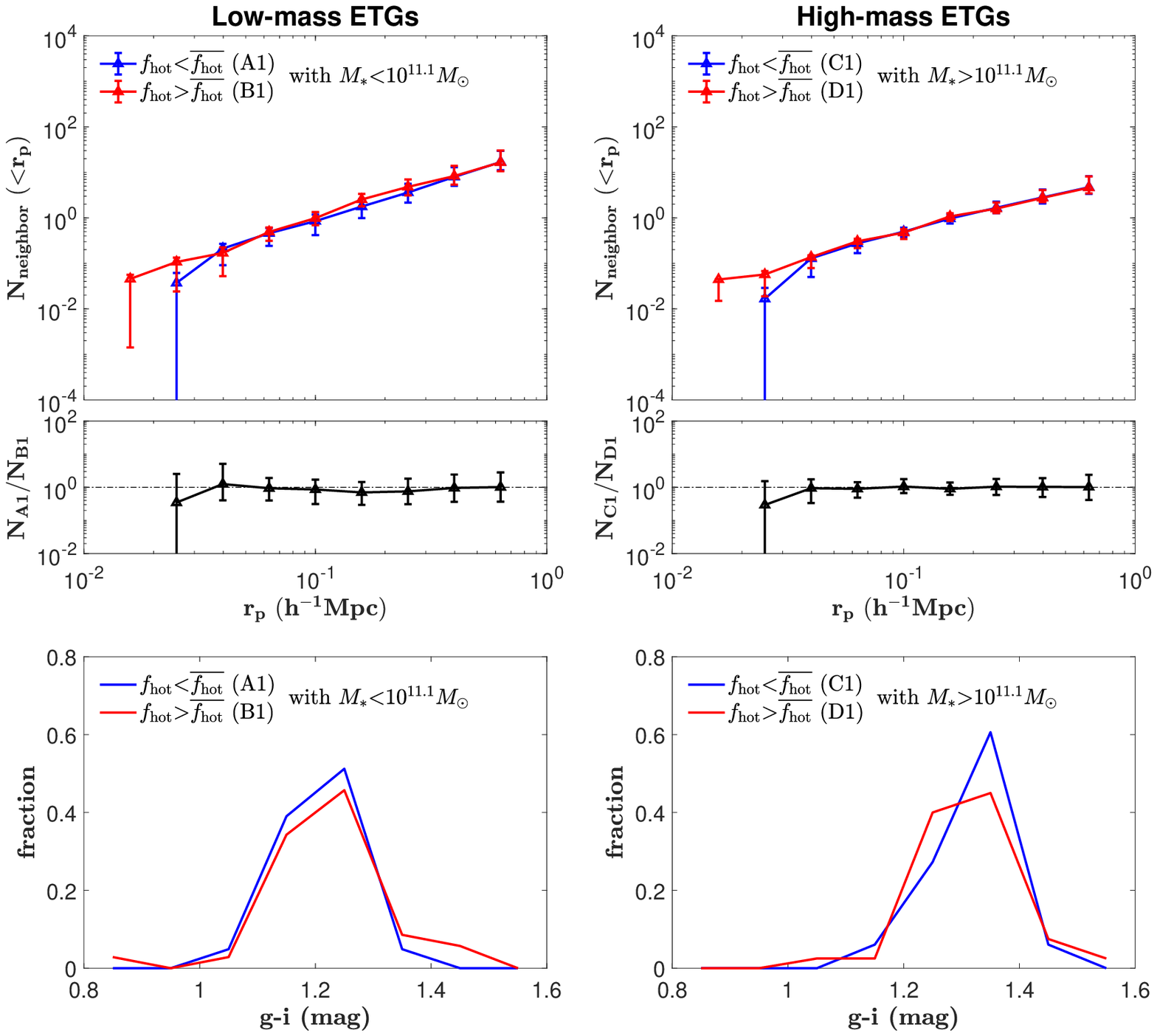}
    \caption{The neighbour counts $N_{\rm neighbour}(\textless r_p)$ as a function of projected radius $r_p$ and the colour distribution of $g-i$ for four different groups of galaxies separated as follows: A1, $M_*\textless10^{11.1}M_{\odot}$ and $f_{\rm hot}\textless\overline{f_{\rm hot}}$; B1, $M_*\textless10^{11.1}M_{\odot}$ and $f_{\rm hot}\textgreater\overline{f_{\rm hot}}$; C1, $M_*\textgreater10^{11.1}M_{\odot}$ and $f_{\rm hot}\textless\overline{f_{\rm hot}}$; and D1, $M_*\textgreater10^{11.1}M_{\odot}$ and $f_{\rm hot}\textgreater\overline{f_{\rm hot}}$. We compare A1 and B1 as a pair on the left and C1 and D1 as another pair on the right panels. The upper panels show the variation of $N_{\rm neighbour}(\textless r_p)$ with $r_p$ for each group of galaxies. In the middle, the black lines with error bars represent the ratios of neighbour counts between two groups in each pair. The bottom panels show the the corresponding normalized $g-i$ colour distribution.}
    \label{environment-fhot}
\end{centering}
\end{figure*}

\begin{figure*}
\begin{centering}
	\includegraphics[width=16cm]{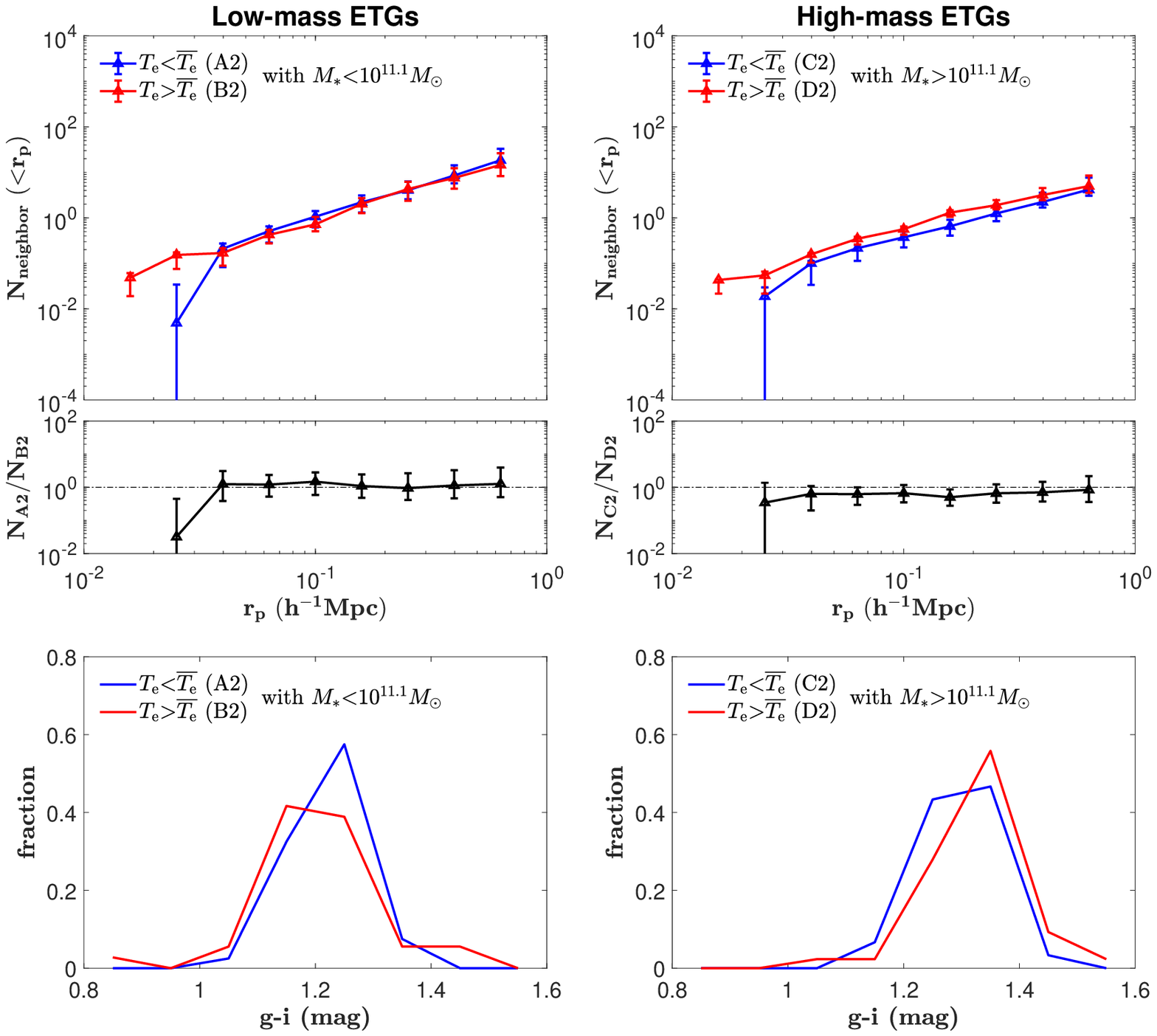}
    \caption{Similarly to Fig.~\ref{environment-fhot}, but four groups separated based on triaxial parameter $T_{\rm e}=\overline{T_{\rm e}}$: A2, $M_*\textless10^{11.1}M_{\odot}$ and $T_{\rm e}\textless\overline{T_{\rm e}}$; B2, $M_*\textless10^{11.1}M_{\odot}$ and $T_{\rm e}\textgreater\overline{T_{\rm e}}$; C2, $M_*\textgreater10^{11.1}M_{\odot}$ and $T_{\rm e}\textless\overline{T_{\rm e}}$; and D2, $M_*\textgreater10^{11.1}M_{\odot}$ and $T_{\rm e}\textgreater\overline{T_{\rm e}}$.}
    \label{environment-T}
\end{centering}
\end{figure*}

As indicated in section $\S$~\ref{sec2}, we investigate the environmental influence on internal galaxy properties by introducing galaxy neighbour counts $N_{\rm neighbour}(\textless r_p)$, which gives the number of galaxies around a galaxy within a projected radius $r_p$ \citep{Li2008}, using a complete photometric reference sample of nearby star-forming galaxies constructed from the New York University Value-Added Galaxy Catalog (NYU-VAGC, \citealp{Blanton2005b}). Neighbours are counted in the vicinity of our galaxies and a statistical correction is applied to remove the effect of chance projections by subtracting the average counts around randomly placed galaxies \citep{Li2008}. In this section, we study the variation of $N_{\rm neighbour}(\textless r_p)$ for galaxies separated into different groups based on two of their internal properties namely the hot orbit fraction $f_{\rm hot}$ and the triaxial parameter $T_{\rm e}$ at effective radius $R_{\rm e}$.

Due to the rapid change of intrinsic structures at $M_*\sim 10^{11.1}M_{\odot}$, we first separate galaxies into low-mass ($M_*\textless10^{11.1}M_{\odot}$) and high-mass ($M_*\textgreater10^{11.1}M_{\odot}$) ETGs. Then for low-mass and high-mass ETGs separately, we perform a linear fit of $f_{\rm hot}$ as a function of stellar mass. The fitted values roughly represent the average $\overline{f_{\rm hot}}$ at different stellar mass $\log(M_*/M_{\odot})$. The low-mass ETGs ($M_*\textless10^{11.1}M_{\odot}$) are then divided into two groups: (1) A1, $f_{\rm hot}\textless\overline{f_{\rm hot}}$; (2) B1, $f_{\rm hot}\textgreater\overline{f_{\rm hot}}$. Similarly for high-mass ETGs ($M_*\textgreater10^{11.1}M_{\odot}$), we also have two groups: (1) C1, $f_{\rm hot}\textless\overline{f_{\rm hot}}$; (2) D1, $f_{\rm hot}\textgreater\overline{f_{\rm hot}}$. A1 and B1 will be a comparison pair with, their major difference being the hot orbit fraction $f_{\rm hot}$ for similar stellar mass. Similarly C1 and D1 will be another pair.

We calculate the neighbour counts for each group of galaxies $N_{\rm neighbour}(\textless r_p)$ as a function of projected radius $r_p$, using SDSS galaxies with r-band magnitude $M_r\textless-20$ as the photometric reference sample.
We perform the calculation several times using bootstrapping, and the variations from the bootstrapping results are used as errors.
In Fig.~\ref{environment-fhot}, we show the comparison between A1 and B1 on the left, and C1 and D1 on the right.
The top panels show the variation of $N_{\rm neighbour}(\textless r_p)$ as a function of $r_p$ for different groups, the middle panels represent the ratios of $N_{\rm neighbour}$ between the corresponding two groups, while the bottom panels are their $g-i$ colour distributions.
A1 and B1 have similar stellar mass and $g-i$ colour distribution, while the one with a higher hot orbit fraction (B1) has a higher $N_{\rm neighbour}$ at $r_p\sim40$ kpc.
Similarly, C1 and D1 have similar stellar mass and $g-i$ colour distribution, and the one with a higher hot orbit fraction (D1) has a higher $N_{\rm neighbour}$ at $r_p\sim40$ kpc.
The corresponding ratios are $N_{\rm A1}/N_{\rm B1}\sim0.3$ and $N_{\rm C1}/N_{\rm D1}\sim0.3$, although there are still relatively large uncertainties.

A similar separation is performed based on the triaxiality $T_{\rm e}$.
We estimate the average triaxiality $\overline{T_{\rm e}}$ as a function of stellar mass $\log(M_{*}/M_{\odot})$ for low-mass and high-mass ETGs separately.
We then divide low-mass ETGs ($M_*\textless10^{11.1}M_{\odot}$) into two groups: (1) A2, $T_{\rm e}\textless\overline{T_{\rm e}}$; (2) B2, $T_{\rm e}\textgreater\overline{T_{\rm e}}$; and similarly divide high-mass ETGs ($M_*\textgreater10^{11.1}M_{\odot}$) into two groups: (1) C2, $T_{\rm e}\textless\overline{T_{\rm e}}$; (2) D2, $T_{\rm e}\textgreater\overline{T_{\rm e}}$.
A2 and B2 will be a comparison pair and, their major difference is the triaxiality $T_{\rm e}$ for similar stellar mass.
Similarly C2 and D2 will be another pair.

We calculate $N_{\rm neighbour}(\textless r_p)$ as a function of projected radius $r_p$ for different groups as shown in Fig.~\ref{environment-T}. A2 and B2 have similar mass and $g-i$ colour distribution. The one with higher $T_{\rm e}$ (B2) has higher $N_{\rm neighbour}$ at $r_p\sim40$ kpc with the ratio $N_{\rm A2}/N_{\rm B2}\sim0.03$. C2 and D2 have similar mass and $g-i$ colour distribution. We also find that the one with higher $T_{\rm e}$ (D2) has higher $N_{\rm neighbour}$ at $r_p\sim40$ kpc. With the ratio $N_{\rm C2}/N_{\rm D2}\sim0.3$, the difference within this pair is not as significant as the pair of A2 and B2.

The above analysis shows that galaxies with more close neighbours ($r_p\sim40$ kpc) tend to be more prolate-like and have more hot orbits. For similar stellar mass and colour distributions for each pair, the differences in the internal structures are unlikely to be caused by their star formation histories. It indicates that close neighbours may affect the internal structures (even in the inner 1 $R_{\rm e}$ regions) of ETGs. Although our sample is relatively small, and the uncertainties are large, we see that the comparison of the four pairs (A1 vs B1, C1 vs D1, A2 vs B2 and C2 vs D2) lead to coherent results.

Our model uncertainties could contribute part of the variations on $T_{\rm e}$ and $f_{\rm hot}$, which could cause the mixture of galaxies in the two groups, separated by an average value of $T_{\rm e}$ and $f_{\rm hot}$ in each pair, thus diluting the differences. The intrinsic differences of $N_{\rm neighbour}$ for the two groups of galaxies in each pair could be larger.

\section{Discussion}
\label{sec8}
Fig.~\ref{DMfrac-vs-mass} shows that our dark matter fractions within one effective radius $f_{\rm DM}(\textless R_{\rm e})=M_{\rm dark}/M_{\rm tot}$ obtained from Schwarzschild modelling are $f_{\rm DM}(\textless R_{\rm e})\sim0.4$ on average for massive ETGs, which is typically higher than the JAM results \citep{Cappellari2013a}. Most massive ETGs are slow rotators and are usually not oblate-like, so the assumptions required for oblate JAM modelling are not met for the dynamical modelling of these galaxies, which means there could be some unphysical biases. Our triaxial Schwarzschild modelling is much more free and does not have this problem. According to our model tests with mock data created from the Illustris simulations \citep{Jin2019}, when the estimations of total masses are accurate, $f_{\rm DM}(\textless R_{\rm e})$ could be overestimated by a factor of $\sim38$ percent if the galaxies have cored dark matter haloes and we model assuming the galaxies should follow the NFW profile. When we use a generalized NFW (gNFW, \citealp{Cappellari2013a}) halo instead, the overestimation can be reduced to $\sim18$ percent. For real MaNGA galaxies, we do not know the true dark matter profiles, and so assume NFW haloes in our models. Thus $f_{\rm DM}(\textless R_{\rm e})$ could be overestimated with an upper limit of $\sim38$ percent. \citet{Cappellari2013a} used both fixed and free NFW haloes in their JAM modelling, and they gave lower $f_{\rm DM}(\textless R_{\rm e})$ values than ours for massive galaxies on average. If the true dark matter profiles of these massive ETGs profile are cored, we could have overestimated the dark matter fractions, and the results from JAM might then be less biased. Although this inconsistency between the two different modelling methods exists, our main conclusions about dark matter fractions and orbital structures are not affected.

In $\S$~\ref{sec6.2} and $\S$~\ref{sec6.3}, we find a rapid change around $\log(M_*/M_{\odot})\sim11.1$ which divides galaxies into two categories. Low-mass ETGs with $\log(M_*/M_{\odot})\textless 11.1$ tend to be oblate-like and dominated by rotation about the minor axis, while high-mass ETGs with $\log(M_*/M_{\odot})\textgreater 11.1$ tend to be prolate-like and dominated by rotation about the major axis and centrophilic orbits. A similar rapid change also exists in Fig.~\ref{DMfrac-vs-mass}. These trends in the inner parts of ETGs are generally consistent with the two formation paths of ETGs proposed by \citet{Cappellari2016}. Slow rotators grow via mergers, are triaxial and dominate the massive region. By comparison, low-mass fast rotators grow via gas accretion and their structure parallels that of spiral galaxies. The increasing fractions of long-axis tube orbits with stellar mass support the scenario that the most massive slow-rotators could form via major dry mergers \citep{Li2018}. Detailed comparison of the orbit distributions between these MaNGA galaxies and cosmologically simulated galaxies could possibly lead to a more quantitative understanding of the formation history.

As alluded to in $\S$~\ref{sec1}, the orbit circularity distributions in $\S$~\ref{sec6.4} being based on only partially constrained, 6-dimensional phase space model data must not be taken as a true representation of a real galaxy. The distributions may be taken however as illustrative or indicative of what the structures might look. In addition, it must be remembered that Schwarzschild's method can only weight the orbits it is given. Changing the given orbits will change the orbits selected by the modelling process, and have an impact on the orbit circularity distribution.

We find no difference on mass distributions, intrinsic stellar shapes and internal orbital structures for central and satellite ETGs in the same mass range. This result is consistent with \citet{Greene2018}, who used two-dimensional kinematics to investigate the difference between central and satellite ETGs. The internal structures of galaxies are dominated by the rapid physical processes associated with the growth of stellar mass. The physical processes due to the difference between centrals and satellites are unlikely to affect galaxy structures, especially in the inner regions ($\textless R_{\rm e}$). However, when considering the local density environments indicated by neighbour counts, we find that galaxies that have higher neighbour counts within $\sim40$ kpc tend to be more prolate-like and have more hot orbits. There is evidence that prolate galaxies are likely originated from major mergers \citep{Tsatsi2017,Li2018}. The increasing fractions of both prograde and counter-rotating long-axis tube orbits with stellar mass as shown in Fig.~\ref{lambda-x} also support this statement. There is no clear evidence whether or not minor mergers or tidal forces from nearby galaxies can affect the stellar orbit distributions of massive ETGs in the inner regions ($\textless R_{\rm e}$). Our results suggest that the $N_{\rm neighbour}$ might be an indicator of galaxies' merger histories. A galaxy with more close neighbours today indicates a denser environment, and this galaxy may have had a higher frequency of major mergers in the past.

All the analysis we have performed has been undertaken using SPS stellar masses to analyse galaxy features or properties determined from dynamical modelling. We therefore have had to deal with analysis using data from different modelling regimes. Using SPS stellar masses where there is no other option, for example in utilising existing MaNGA or SDSS catalogues, is understandable. Dynamical masses, both stellar and dark matter, come as a pair in the sense that there is a degeneracy between them that is not resolvable with the galaxy observations that are currently available. Both \citet{Li2016} and \citet{Jin2019} make the point however that the total mass is recovered well. Thus, there is is no good reason to discard dynamical stellar masses for SPS masses when calculating dark matter fractions. Similarly, orbits and values derived from them are related to the total mass in a galaxy. Why researchers choose to ignore total mass and just analyse orbits with respect to stellar mass is unclear, and why an SPS mass is sometimes substituted for a dynamical stellar mass is even less clear. It may be that uncertainties about a galaxy's dark matter distribution are considered to be  ``large" with an associated effect on the stellar distribution. Switching to an SPS stellar mass will not resolve the issues with the dark matter distribution. What we have highlighted here is that cross-overs between modelling regimes needs to be justified and communicated effectively. We will continue to progress this in the future.

\section{Summary}
\label{sec9}
We have met the objectives we set out in $\S$~\ref{sec1} in that we have
\begin{enumerate}
\item modelled the selected MaNGA early-type galaxies individually using Schwarzschild's method and determined their mass distributions,  intrinsic stellar shapes and internal orbit distributions,
\item examined statistically the differences and similarities between central and satellite ETGs, and
\item assessed the role of the environment in the galaxies' evolution.
\end{enumerate}

We set out more detail on our findings below. Significantly we find the intrinsic properties of ETGs have a rapid change at about $\log(M_*/M_{\odot})\sim11.1$.
\begin{itemize}
\item Dark matter fractions within one effective radii $f_{\rm DM}(\textless R_{\rm e})$ increase with total stellar mass for our sample. Low-mass ETGs ($\log(M_*/M_{\odot})\textless11.1$) have on average $f_{\rm DM}(\textless R_{\rm e})\sim0.2$ and high-mass ETGs ($\log(M_*/M_{\odot})\textgreater11.1$) have on average $f_{\rm DM}(\textless R_{\rm e})\sim0.4$.

\item The stellar components of low-mass ETGs ($\log(M_*/M_{\odot})\textless11.1$) tend to be oblate-like, with their average triaxiality $T_{\rm e}\sim0.4$, while these of high-mass ETGs ($\log(M_*/M_{\odot})\textgreater11.1$) tend to be prolate-like, with their average triaxiality $T_{\rm e}\sim0.6$.

\item Low-mass ETGs ($\log(M_*/M_{\odot})\textless11.1$) have more cold and warm orbits ($f_{\rm cold}+f_{\rm warm}+f_{\rm CR}\sim0.55$, $f_{\rm prolong}+f_{\rm CRlong}\sim0.15$ and $f_{\rm box}+f_{\rm SR}\sim0.3$ on average), which means they are dominated by rotation about the minor axis. High-mass ETGs ($\log(M_*/M_{\odot})\textgreater11.1$) have more long-axis tube orbits and box orbits ($f_{\rm cold}+f_{\rm warm}+f_{\rm CR}\sim0.35$, $f_{\rm prolong}+f_{\rm CRlong}\sim0.25$ and $f_{\rm box}+f_{\rm SR}\sim0.4$ on average). The amount of rotation about the major axis, rotation about the minor axis and centrophilic orbits are comparable for high-mass ETGs. The variation of orbital fractions as a function of stellar mass are similar to those ETGs in CALIFA sample, but not quantitatively comparable due to very different selection functions.

\item There are not significant differences in the above property values between central and satellite ETGs for the same stellar masses. Thus, the variation of intrinsic orbital shapes and dark matter fractions are driven by the variation in stellar masses of galaxies. Being centrals or satellites does not have noticeable effects on these properties, at least for the inner regions covered by the MaNGA observations.

\item We find that early-type galaxies more prolate-like or with higher hot orbit fractions tend to have higher close neighbour counts at $r_p\sim40$ kpc, for similar stellar mass and colour distribution. This is consistent with the major merger origin of prolate galaxies. A galaxy has more close neighbours today may indicate a denser environment, thus have had higher major merging frequency in the past.
\end{itemize}

Having summarised our findings above, it must be remembered that all the material involving orbit circularities is at best indicative because of the observational capabilities available and the consequent deprojection degeneracies.

\section*{Acknowledgements}
We thank R. C. E. van den Bosch for providing us his triaxial Schwarzschild software and M. Cappellari for making his MGE software publicly available. The modelling was accomplished on the ``Zen'' cluster at National Astronomical Observatories, Chinese Academy of Sciences (NAOC) and on ``Venus'' at Tsinghua University. This work is partly supported by the National Key Basic Research and Development Program of China (grant number 2018YFA0404501 to SM), by the National Science Foundation of China (grant numbers 11821303, 11333003, 11390372 and 11761131004 to SM). LZ acknowledges support from Shanghai Astronomical Observatory, Chinese Academy of Sciences under grant number Y895201009. GvdV acknowledges funding from the European Research Council (ERC) under the European Union's Horizon 2020 research and innovation programme under grant agreement No 724857 (Consolidator Grant ArcheoDyn).

Funding for the Sloan Digital Sky Survey IV has been provided by the Alfred P. Sloan Foundation, the U.S. Department of Energy Office of Science, and the Participating Institutions. SDSS acknowledges support and resources from the Center for High-Performance Computing at the University of Utah. The SDSS web site is www.sdss.org.

SDSS is managed by the Astrophysical Research Consortium for the Participating Institutions of the SDSS Collaboration including the Brazilian Participation Group, the Carnegie Institution for Science, Carnegie Mellon University, the Chilean Participation Group, the French Participation Group, Harvard-Smithsonian Center for Astrophysics, Instituto de Astrof\'isica de Canarias, The Johns Hopkins University, Kavli Institute for the Physics and Mathematics of the Universe (IPMU) / University of Tokyo, the Korean Participation Group, Lawrence Berkeley National Laboratory, Leibniz Institut f\"ur Astrophysik Potsdam (AIP), Max-Planck-Institut f\"ur Astronomie (MPIA Heidelberg), Max-Planck-Institut f\"ur Astrophysik (MPA Garching), Max-Planck-Institut f\"ur Extraterrestrische Physik (MPE), National Astronomical Observatories of China, New Mexico State University, New York University, University of Notre Dame, Observat\'orio Nacional / MCTI, The Ohio State University, Pennsylvania State University, Shanghai Astronomical Observatory, United Kingdom Participation Group, Universidad Nacional Aut\'onoma de M\'exico, University of Arizona, University of colourado Boulder, University of Oxford, University of Portsmouth, University of Utah, University of Virginia, University of Washington, University of Wisconsin, Vanderbilt University, and Yale University.

\appendix
\section{Comparing the orbit circularity distributions with CALIFA results}
\begin{figure*}
\begin{centering}
	\includegraphics[width=16cm]{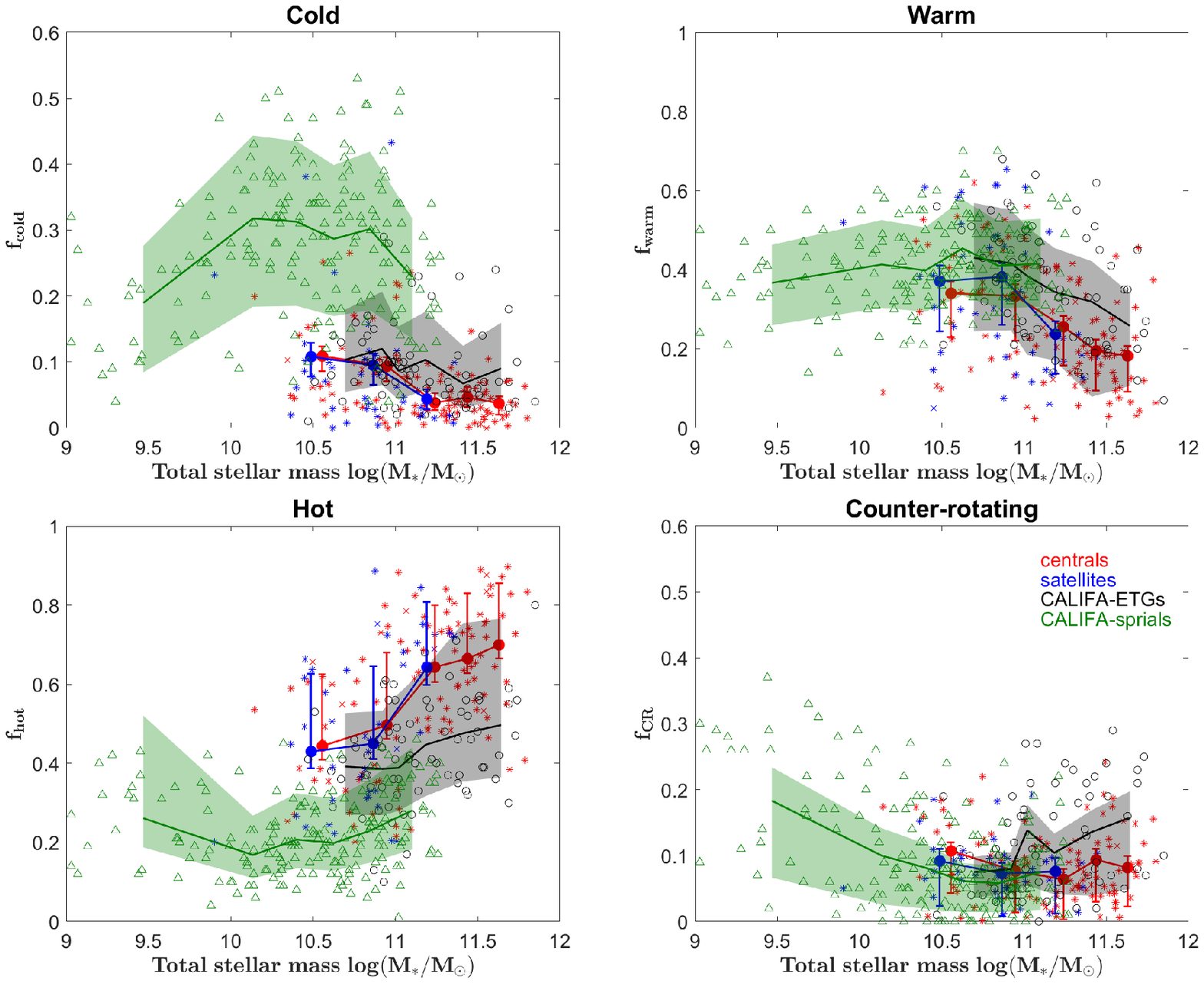}
    \caption{The luminosity fractions of cold, warm, hot and counter-rotating orbits versus total stellar mass, including results from both MaNGA and CALIFA. From top-left to bottom-right, they are: cold ($\lambda_z\ge0.8$), warm($0.25<\lambda_z<0.8$), hot($-0.25\le\lambda_z\le0.25$) and counter-rotating ($\lambda_z<-0.25$) components. The red and blue symbols indicate our MaNGA sample and are exactly the same as Fig.12 in the manuscript. The black circles represent the results of ETGs from Zhu et al. (2018)'s CALIFA sample, while the green triangles represent that of spirals. The corresponding solid lines with shadows show the mean binned values and their uncertainties.}
    \label{compare-vs-CALIFA}
\end{centering}
\end{figure*}
\begin{figure*}
\begin{centering}
	\includegraphics[width=14cm]{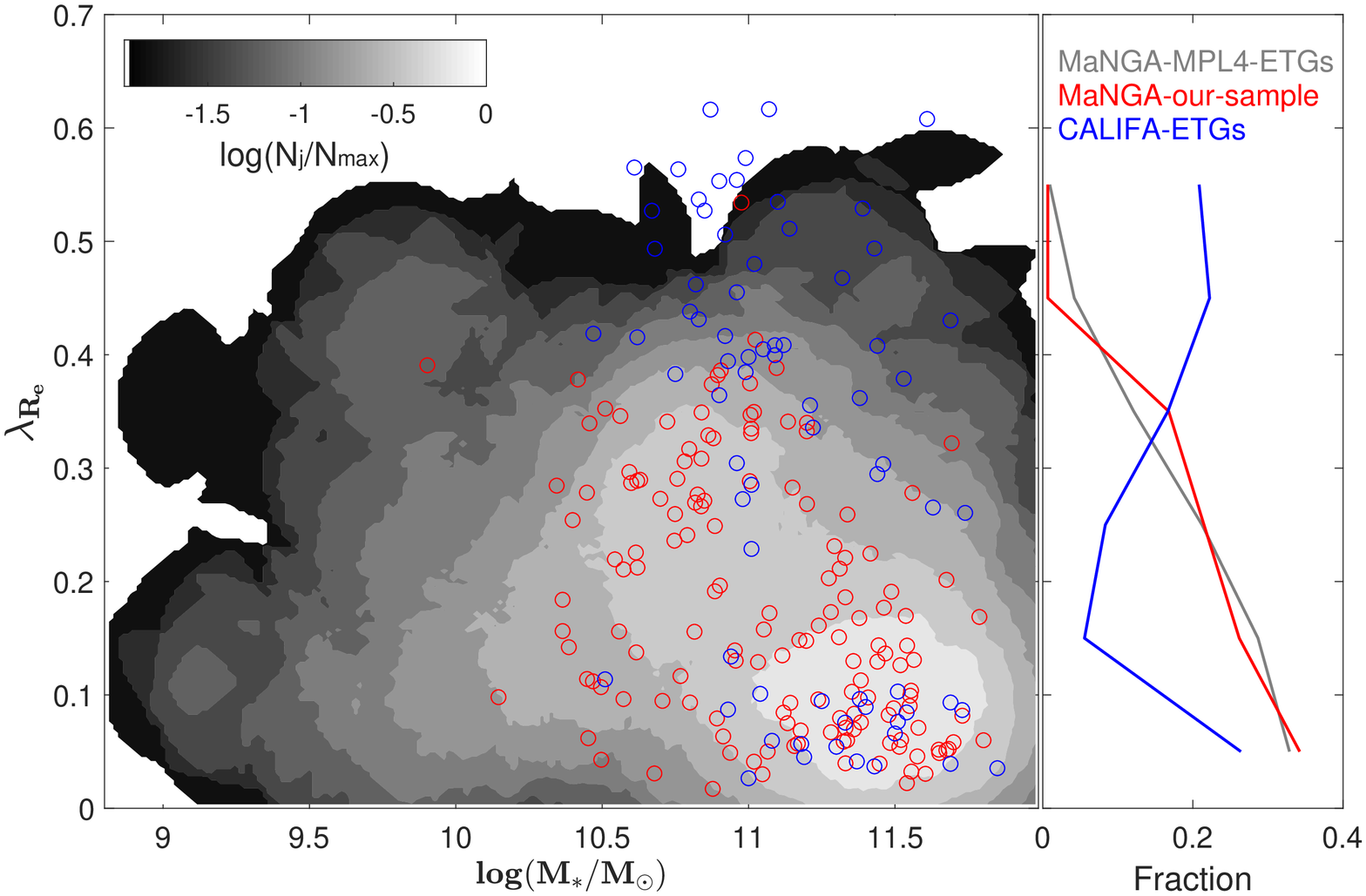}
    \caption{A widely used two-dimensional rotation indicator $\lambda_{R_{\rm e}}=\frac{\langle R\left| V \right|\rangle}{\langle R\sqrt{V^2+\sigma^2}\rangle}$ as a function of total stellar mass $\log(M_*/M_{\odot})$. The red circles represent our MaNGA sample and the blue circles represent the CALIFA ETGs, while the background colour map shows the number count distribution of all ETGs in the MaNGA MPL4, with their number densities $\log(N_j/N_{\rm max})$ being indicated by the colour bar. $N_j$ is the number of galaxies in mass vs $\lambda_{R_{\rm e}}$ bin $j$, and $N_{\rm max}$ is the maximum of the $N_j$. The right panel shows the corresponding marginalised fractions of galaxies by $\lambda_{R_{\rm e}}$. We can see that the distribution of $\lambda_{R_{\rm e}}$ is quite different between our MaNGA sample and CALIFA sample. CALIFA sample have more fast-rotating galaxies than our MaNGA sample, which is consistent with the 3D modelling results. The marginalised fractions of our sample are nearly the same as that of all the ETGs in MaNGA MPL4 (we select our sample from MaNGA MPL4), which means this difference is not caused by our sample selection criteria.}
    \label{lambda_Re}
\end{centering}
\end{figure*}

In Fig.~\ref{compare-vs-CALIFA}, we compare the distribution of $\lambda_z$ with the results in \citet{Zhu2018b}, who applied this triaxial Schwarzschild implementation to 300 CALIFA galaxies. For early-type galaxies, the CALIFA sample has more cold, warm and counter-rotating orbits, and less hot orbits than MaNGA sample on average, which means the CALIFA sample tends to have more rotating galaxies. This difference is mainly caused by the different selection functions between MaNGA and CALIFA. The MaNGA sample is missing a large fraction of fast-rotating ETGs (see Fig.~\ref{lambda_Re}).

\section{Comparison between SPS and dynamical stellar mass}
\begin{figure*}
\begin{centering}
	\includegraphics[width=12cm]{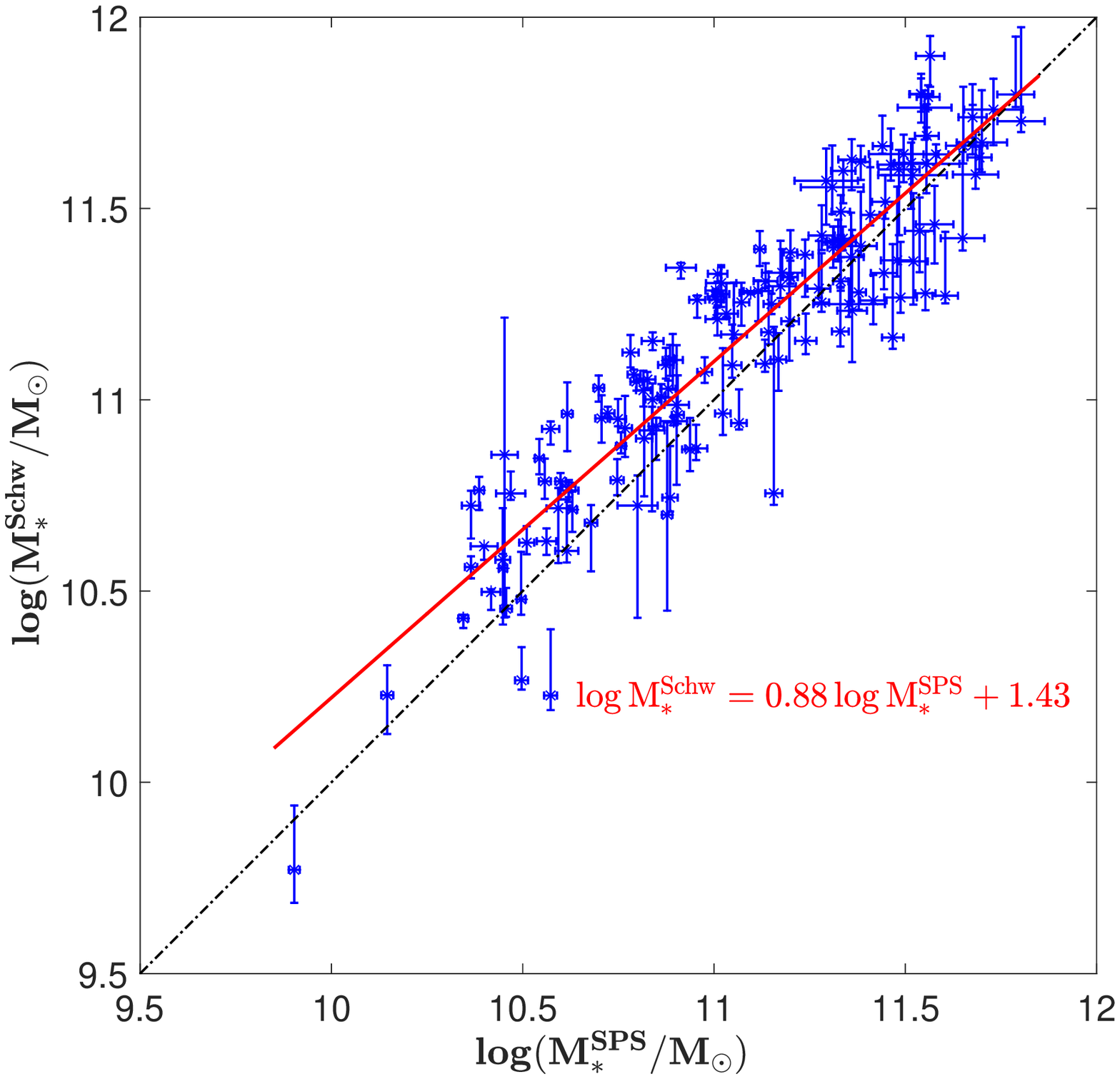}
    \caption{The comparison between stellar mass inferred from stellar population synthesis (SPS) and our Schwarzschild modelling. The horizontal axis is for SPS stellar mass while the vertical axis is for dynamical stellar mass. The best-fitting values in Schwarzschild modelling and SPS are indicated by the blue asterisks. The vertical error bars represent $1\sigma$ confidence levels in Schwarzschild modelling, while the horizontal error bars represent the scatters estimated from \citep{Ge2018}. The black dashed line indicates equal values and the red line is the linear fitted line.}
    \label{mass-comparison}
\end{centering}
\end{figure*}

We present a comparison of the stellar masses determined from stellar population synthesis (SPS) and our Schwarzschild modelling in Figure~\ref{mass-comparison}. The masses, coming from the two different modelling regimes, are clearly different. The SPS stellar masses are calculated without any knowledge of galaxy dark matter haloes, while the Schwarzschild stellar masses are determined concurrently with the dark matter masses and are affected by degeneracies between the two mass types. Fitting a straight line to the  mass values gives $\rm \log M_*^{Schw}=0.88\log M_*^{SPS}+1.43$.

\section{Example data table}
The major properties of ten example galaxies in our sample and the corresponding modelling results are available in Table~\ref{example-data-table}. The full appendix table that contains all our sample galaxies is available as supplementary material on the journal website.

\begin{table*}
\caption{The major properties of ten galaxies in our sample and the corresponding modelling results. From left to right, they are: (1) MaNGA ID; (2) the total stellar mass $\log(M_*/M_{\odot})$ ; (3) the effective radius $R_{\rm e}$; (4) the dark matter fraction within one effective radius $f_{\rm DM}(\textless R_{\rm e})$; (5) the medium to long axis ratio $p_{\rm e}$, the short to long axis ratio $q_{\rm e}$ and the triaxial parameter $T_{\rm e}=(1-p_{\rm e}^2)/(1-q_{\rm e}^2)$ at one effective radius; (6) the fractions of orbits $f_{\rm cold}$, $f_{\rm warm}$, $f_{\rm hot}$ and $f_{\rm CR}$; (7) the fractions of orbits $f_{\rm prolong}$, $f_{\rm CRlong}$, $f_{\rm box}$ and $f_{\rm SR}$. Please see the journal website for the complete table.}
\begin{tabular}{|>{\centering\arraybackslash}p{1.35cm}|>{\centering\arraybackslash}p{1.5cm}|>{\centering\arraybackslash}p{1.06cm}|>{\centering\arraybackslash}p{1.43cm}|*{3}{>{\centering\arraybackslash}p{0.55cm}}|*{4}{>{\centering\arraybackslash}p{0.55cm}}|*{4}{>{\centering\arraybackslash}p{0.6cm}}|}
\hline
\multicolumn{3}{|c|}{Major properties} & \multicolumn{12}{c|}{Schwarzschild modelling results}\\
\hline
MaNGA ID & $\log(M_*/M_{\odot})$ & $R_{\rm e}$ (kpc) & $f_{\rm DM}(\textless R_{\rm e})$ & $p_{\rm e}$ & $q_{\rm e}$ & $T_{\rm e}$ & $f_{\rm cold}$ & $f_{\rm warm}$ & $f_{\rm hot}$ & $f_{\rm CR}$ & $f_{\rm prolong}$ & $f_{\rm CRlong}$ & $f_{\rm box}$ & $f_{\rm SR}$\\
\hline
1-259250 & 10.89 & 1.82 & 0.063 & 0.937 & 0.622 & 0.198 & 0.086 & 0.527 & 0.352 & 0.035 & 0.032 & 0.051 & 0.179 & 0.090 \\
1-285095 & 11.50 & 15.94 & 0.754 & 0.980 & 0.697 & 0.077 & 0.045 & 0.182 & 0.535 & 0.242 & 0.001 & 0.023 & 0.309 & 0.201 \\
1-285066 & 10.91 & 5.34 & 0.120 & 0.849 & 0.739 & 0.616 & 0.015 & 0.180 & 0.730 & 0.071 & 0.242 & 0.081 & 0.244 & 0.163 \\
1-46825 & 11.18 & 7.38 & 0.498 & 0.858 & 0.779 & 0.670 & 0.033 & 0.123 & 0.770 & 0.077 & 0.155 & 0.105 & 0.276 & 0.235 \\
1-564264 & 11.48 & 10.08 & 0.327 & 0.975 & 0.916 & 0.309 & 0.008 & 0.238 & 0.560 & 0.193 & 0.133 & 0.089 & 0.240 & 0.098 \\
1-605419 & 11.69 & 8.58 & 0.001 & 0.997 & 0.946 & 0.057 & 0.072 & 0.456 & 0.384 & 0.089 & 0.050 & 0.047 & 0.075 & 0.211 \\
1-585593 & 10.86 & 2.10 & 0.087 & 0.939 & 0.517 & 0.162 & 0.166 & 0.472 & 0.271 & 0.092 & 0.018 & 0.029 & 0.132 & 0.090 \\
1-156062 & 11.31 & 12.00 & 0.095 & 0.822 & 0.789 & 0.859 & 0.018 & 0.106 & 0.840 & 0.033 & 0.199 & 0.231 & 0.210 & 0.201 \\
1-235530 & 10.56 & 1.66 & 0.034 & 0.934 & 0.893 & 0.624 & 0.042 & 0.267 & 0.619 & 0.072 & 0.156 & 0.239 & 0.158 & 0.064 \\
1-320584 & 11.68 & 8.92 & 0.440 & 0.913 & 0.754 & 0.387 & 0.071 & 0.305 & 0.550 & 0.074 & 0.109 & 0.079 & 0.292 & 0.071 \\

\hline
\end{tabular}
\label{example-data-table}
\end{table*}

\end{document}